# Global Abiotic Sulfur Cycling on Earth-like Terrestrial Planets

Rafael Rianço-Silva[1, 2, 3, 5] *, Javed Akhter Mondal[4, 5] *, Matthew A. Pasek[6], Henry Jurney[5], Marcos Jusino-Maldonado[5, 7], Henderson James Cleaves II[5, 8, 9, 10]

**Affiliations**

1. Instituto de Astrofísica e Ciências do Espaço, Observatório Astronómico de Lisboa, Ed. Leste, Tapada da Ajuda, 1349-018 Lisbon, Portugal
2. Departamento de Física, Faculdade de Ciências, Universidade de Lisboa, 1749-016 Lisboa, Portugal
3. Department of Physics and Astronomy, University College London, Gower Street, WC1E 6BT London, United Kingdom
4. Deep Space Exploration Laboratory/CAS Key Laboratory of Crust-Mantle Materials and Environments, University of Science and Technology of China, Hefei 230026, China
5. Blue Marble Space Institute of Science, Seattle, USA
6. Earth and Environmental Science, Rensselaer Polytechnic Institute, Troy NY, USA
7. Planetary Habitability Laboratory, University of Puerto Rico at Arecibo, Puerto Rico
8. Earth-Life Science Institute, Tokyo Institute of Technology, Tokyo, Japan
9. Earth and Planets Laboratory, Carnegie Institution of Washington, Washington DC, USA
10. Department of Chemistry, Howard University, Washington DC, USA

* Authors contributed equally for this work

Corresponding author email: rdsilva@fc.ul.pt

**Abstract**

Sulfur is a redox active element that may have helped mediate an electron flow that kickstarted life and which presently is an essential element for all life on Earth. Despite current uncertainties in global sulfur fluxes, modeling sulfur's abiotic cycling through Earth's deep history is important for understanding the impact of a planet-wide biosphere on sulfur's geochemical cycling and availability and *vice versa*. We present here an open-source, dynamical box model for estimating global sulfur fluxes and concentrations among surface and deep Earth reservoirs over Earth history, allowing tracking and estimation of the sulfur distribution in planetary reservoirs over deep time in the absence of life. While the main model presented here does not take into account the abrupt evolution of redox-shunting biosynthetic pathways such as oxygenic photosynthesis, we also modeled the abiotic sulfur cycle before and after a Great Oxidation Event (GOE)-like transition on Earth-like planets. Our results suggest a considerably distinct chemical makeup of sulfur content in marine sediments in the absence of life on an Earth-like planet, leading to a marine sediment sulfate content two orders of magnitude larger than on present-day Earth and a marine sediment sulfide content 4 orders of magnitude lower than on present-day Earth, attributable to the lack of microbial sulfur metabolism. This model could be useful for understanding sulfur cycling on potentially habitable exoplanets.

1. Introduction

The distribution of redox-sensitive volatile and semi-volatile elements on planets may provide important interpretable information about planetary evolution (*e.g.* Schindler & Kasting, 2000). Cycles of carbon, nitrogen, sulfur, oxygen and hydrogen, among other elements, are presently heavily influenced by life on Earth (see for example Falkowski, 1997). In the absence of life, these elements might be distributed differently on Earth between its atmospheric, oceanic and lithospheric reservoirs, and this distribution could serve as a biosignature for remote life detection on extrasolar planets (Krissansen-Totton *et al.*, 2018; Schwieterman *et al.*, 2018; Madhusudhan *et al.*, 2025).

Sulfur is often overlooked in understanding the origins of life and the evolution of the biosphere, though it figures prominently in some models for these processes (*e.g.* Urey, 1952; Wachtershauser, 1988; Heinen & Lauwers, 1996; De Duve, 1998; Hartman, 1998; Goldford *et al.*, 2017; Olson, 2021). Sulfur's common oxidation states range from +6 for sulfate to -2 for sulfide, presenting an eight electron redox range spanning various speciations of widely differing chemical reactivity. This gives sulfur an eightfold larger ability to move electrons relative to iron on a per atom basis, which accounts for 5.8-6.5% of the bulk silicate Earth (BSE) by mass (Lyubetskaya & Korenaga, 2007), but only cycles one electron between +2 and +3 oxidation states. Thus, sulfur which constitutes only 0.02% by mass of the BSE (200-250 ppm, Kiseeva, 2021), assuming complete redox change, may only shuttle ~1/30$^{th}$ the number of electrons as Fe, but do so in a more kinetically facile fashion.

Sulfur appears in the periodic table immediately below oxygen (O), and is thus less electronegative, undergoing redox reactions more easily and reversibly than O. While O almost always appears in the -2 oxidation state in solids, sulfur has a wider range of natural oxidation states. S metabolism is not only crucial for all extant biochemistry, but sulfur biogeochemistry plays a major role in global element cycling (Fike *et al.*, 2015). Bacterial biomass is ~1 % sulfur by dry weight (Neidhardt *et al.*, 1990). Among the major chemical elements used in biochemistry, sulfur plays an important role as a redox mediator, and is a significant component of biological material in the form of the amino acids cysteine and methionine (Bin *et al.*, 2017), as well as biological cofactors including CoA, thiamine, and Fe-S clusters (Francioso *et al.,* 2020).

*1.1 - Earth's Modern Sulfur Cycle*

Earth's modern sulfur cycle is one of the largest global biogeochemical cycles in terms of atom and electron flux, and is heavily influenced by biology (Fike *et al.*, 2015). The modern biogeochemical sulfur cycle likely evolved alongside biology: the evolution of oxygenic photosynthesis both created a photoprotective ozone layer which changed the rates of photochemical processing of atmospheric and surface sulfur species, and greatly enhanced the rate of sulfur oxidation in the oceans and weatherable mineral surfaces (Stüeken *et al.*, 2012). The rise of biogenic $O_2$ further enabled the widespread metabolism of sulfate reducing microbes, which are estimated to presently turn over some $3.6 \times 10^{11}$ kg S per year globally (Wasmund *et al.*, 2017). Dissimilatory sulfur reduction may have already evolved by the time of the first evidence of life on Earth (Ueno *et al.*, 2008), significantly predating the Great Oxidation Event (GOE), which is discussed below, and underscores the ancient influence of biology on global sulfur cycling.

Sulfur is produced via the α-process in stars by the fusion of Si and He nuclei, and partitions in larger planetary bodies according to its affinity for elements including transition metals (Truran & Heger, 2010). Sulfur is classified as a volatile element astronomically and as a chalcophile element according to the Goldschmidt classification (Goldschmidt, 1937), meaning that rocky planets may be depleted in sulfur due to their formation histories in inner solar system environments, via volatile loss, and still be surface-enriched in sulfur due to planetary dynamics. However, a study by Dasgupta *et al.* (2009) suggests that sulfur's partitioning is not exclusively chalcophile; rather it strongly dependent on redox state and coexisting phases, and under reducing conditions sulfur can partition efficiently into Fe-rich metallic/sulfide melts during core formation.

The planetary chemodynamic behavior of sulfur is especially interesting due to sulfur's relatively facile redox changes, and the markedly different chemodynamic behavior of its different oxidation state species due to their pH- and temperature-dependent solubilities in water, vapor pressures, and other chemical idiosyncrasies. As an example of these marked differences in chemodynamic behavior, sulfide forms extremely insoluble complexes with transition metals. The $K_{sp}$ of iron (II) sulfide is ~$1.4 \times 10^{-17}$ at standard temperature and pressure (STP) (Morse *et al.*, 1987), whereas sulfate salts are orders of magnitude more soluble (*e.g.* the solubility of $CaSO_4$ as the dihydrate (gypsum) is ~15 mM at STP (Meijer & Van Rosmalen, 1984)). Thus geologically rapid and biologically-mediated redox reactions of sulfur likely greatly altered sulfur's solubility-related chemodynamics.

The partitioning of sulfur species between aqueous and gaseous phases is important in considering sulfur's behavior in geochemical reservoirs. The Henry's Law constants for sulfide, sulfite and sulfate species between gas and aqueous phases vary widely (Sander, 2015). As sulfide, sulfite and sulfate exist in equilibrium as various protonated and unprotonated species, their Henry's Law constants determine how they partition according to the pH of the underlying aqueous solution. Thus the acidity of the early oceans may have been important for sulfur chemodynamics (Halevy & Bachan, 2017).

The differing aqueous solubilities of different redox state sulfur species make sulfur cycling on rocky planetary bodies dependent on surface temperatures in especially complex ways. For example, on planets or moons receiving relatively less stellar insolation (for example Titan (Flasar *et al.*, 2005)), volatile sulfur species may also be inhibited from engaging in global geochemical cycles, as the freezing points of both $H_2S$ and $SO_2$ are relatively low (-85.5° and -79° C at STP, respectively). Cold planetary bodies, those with surface temperatures below the freezing point of water, will likely have simple sulfur cycles. On this note, Snowball Earth episodes, such as the ones thought to have occurred at the Cryogenian period of the Neoproterozoic (720 Ma to 640 Ma before present, Hoffman *et al.*, 2017) likely resulted in major perturbations to the sulfur cycle. Isotopic data attests to a drastic drawdown of marine sulfate from the subglacial ocean by a factor of two after the onset of the global glaciation, possibly caused by the decrease in the input of continental sulfur erosion and weathering delivery of sulfur into the oceans and a relative increase of pyrite burial in marine sediments associated with microbial sulfate reduction (Sansjofre *et al.*, 2016).

### *1.2 - Common Solid Earth Geochemical Sulfur Speciation*

Common solid Earth geochemical forms of sulfur include reduced mineral species such as metal sulfides, most importantly iron sulfides (pyrite, *etc.*), as well as oxidized species such as gypsum, anhydrite and barite ($BaSO_4$). Major dissolved sulfur species include sulfate ($SO_4^{2-}$) and hydrosulfide ($SH^-$). In the modern terrestrial atmosphere, the major sulfur species are carbonyl sulfide (OCS with a mixing ratio of 0.5 ppb), $SO_2$ (~20 ppt to >1 ppb) and $H_2S$ (10-200 ppt), along with minor amounts of dimethyl sulfide, $(CH_3)_2S$, (<200 ppt), and carbon disulfide ($CS_2$), with <2 ppt (Tyndall & Ravishankara, 1991; Watts, 2000; Adame *et al.*, 2020). The atmospheric lifetimes of these species in the modern terrestrial atmosphere are on the order of months to years (Tyndall & Ravishankara, 1991), thus it is unlikely they could be maintained out of equilibrium for long without significant volcanic or biological input. OCS, despite being the most abundant S-bearing compound in the modern terrestrial atmosphere due to its long photochemical lifetime (> 1 year), is mostly biogenic, originating from the degradation of dissolved organic matter in the Earth's oceans and fossil fuel burning (Watts, 2000; Lee *et al.*, 2016). Chin & Davis (1993) estimate volcanic outgas of OCS amounts for 1.5% of the total OCS input to modern atmosphere. Recent studies have also pointed to the possibility of H2S contributing to the production of organic hazes in Archean Earth's conditions (Read *et al*, 2022).

Presently, global transport of sulfur is largely determined by geological processes, though biology significantly perturbs these processes. For example, it is generally believed that Earth underwent a rapid oxygenation process ~2.4 Ga (known as the GOE, Luo *et al.*, 2016; Gumsley *et al.*, 2017) which shifted the predominant form of sulfur in Earth's oceans from reduced to oxidized and altered the photochemical processing of atmospheric sulfur to produce detectable diagnostic isotopic signals (*e.g.* $\delta^{34}S$ isotopic fractionation, Farquhar *et al.,* 2000). A more reduced environment, alongside lower erosion input of $S_{OX}$ into the oceans due to smaller continents (Armstrong, 1981) helps explain the lack of major calcium sulfate deposits before the GOE, though scattered barite deposits are observed in the geological record (Jewell, 2000). There is little evidence for enhanced sulfate deposition in the early Earth geological record, but this could track with the lack of evidence for early carbonate minerals, *i.e.*, Earth's dynamic chemodynamic evolutionary processes have removed them. It should be noted it has been suggested that biological sulfate reduction arose before the GOE (Shen *et al.*, 2001; Ueno *et al.,* 2008).

### *1.3 - Chemodynamic Modeling of Sulfur Cycling*

As an extension of our previous work modeling the evolution of the N and P cycles in the absence of biology (Laneuville *et al.,* 2018; Jusino-Maldonado *et al.,* 2022), we here report a general modifiable model for the evolution of sulfur cycling on Earth-like planets on which life has not arisen, for the sake of comparison with "living" planets like Earth, which may be useful for detecting remote biosignatures (Seager, 2014). Many geochemical cycles are interconnected though some may be more affected by biological intervention than others, and the extent to which any given element can be a marker of biological intervention allows many chemical elements to be potential biomarkers in context. The size of an elemental planetary reservoir of a given element and the degree to which it can be perturbed by biology contributes to its utility as a remote biomarker. Sulfur was the next logical element after nitrogen and phosphorus to explore in this context, though the complexity of sulfur's behavior led us to explore it last.

Biology connects multiple elements during geochemical cycling, for example via phenomena such as the Redfield ratio, in which the abundance of more than one element determines biological productivity (Redfield, 1958). Any global model of element cycling on modern Earth is prone to error due to exclusion of as yet unknown processes, thus exploration of historic elemental chemodynamics in the absence of unknown processes is likely prone to at least similar uncertainty. Nevertheless baseline estimates such as the one presented here can be useful for understanding planetary elemental cycling processes. It is worth emphasizing that although the model presented here is based on a lifeless Earth-like planet, considering the very distinct geochemistries of a pre-GOE and a post-GOE Earth (Farquhar *et al.,* 2000) and the expected stark differences of their abiotic sulfur cycles as end-member scenarios (Reinhard *et al.,* 2009; Stueken *et al.,* 2012; Wang *et al.,* 2019), we present in this work two distinct model configurations, as described in section 2.2. One of these considers abiotic pre-GOE Earth S cycling, while the other considers post-GOE continental weathering processes, but still excludes biological processes such as microbial sulfate reduction. The models provided here can be used to evaluate the evolution of sulfur cycling on terrestrial exoplanets and the interpretation of measurable abundances of sulfur species as biosignatures on exoplanets.

### *1.4 - Earth's Sulfur Inventory*

Earth's initial and global sulfur inventory can be estimated by the measured sulfur concentration in meteorites (*e.g.* Gibson *et al.,* 1985). Simple calculations suggest the majority of Earth's primordial sulfur budget is sequestered in its core (*e.g.* Dreibus & Palme, 1996). The BSE and surface sulfur inventory is presumed to be what was left over after Earth's early differentiation, *e.g.* by sulfur segregation and partitioning in deep melts (Wedepohl, 1984).

Compared to our previous explorations of nitrogen and phosphorus cycling (Laneuville *et al.,* 2018; Jusino-Maldonado *et al.,* 2022), sulfur is considerably more complicated, both due to its range of oxidation states and its high BSE abundance. Sulfur constitutes an estimated $3.5 \times 10^{-2}$ by mass % of the BSE (Haynes, 2016), a value intermediate between those of P ($\sim 1.05 \times 10^{-1}$ %, Haynes, 2016) and N ($\sim 1.9 \times 10^{-3}$ %, Haynes, 2016). However S, N and P cycle via markedly different dynamics. As a further example of the importance and nuance of sulfur's chemodynamic history, it has been estimated that biological processes have exported $\sim 1.4 \times 10^{22}$ moles of $H_2$ into the atmosphere over Earth history (most of which has been lost by escape to space, Pope *et al.*, 2012)), representing 2.8 x $10^{22}$ moles of electrons, while biogeochemical cycling of S between the -2 and +6 oxidation states turns over 1.13 x $10^{13}$ moles of electrons per year (Wasmund *et al.*, 2017), roughly 4.5 x $10^{21}$ moles of electrons if continuous over 4 Ga, roughly half the number of electrons lost via $H_2$. Thus sulfur cycling may be comparable in importance to the development of oxygenic photosynthesis in generating a remotely detectable biosignature.

The chemodynamics of sulfur are both intrinsic, due to the chemical nature of the various redox states of sulfur species, and dependent on planetary idiosyncrasies such as the surface temperature of the planet, and atmospheric, oceanic and tectonic dynamics. Examination of the estimated sulfur contents of Venus, Earth and Mars reveals significant differences (see Table 1), some of which can clearly be attributed to Earth's hydrologic cycle, which is partly enabled by Earth's mass and orbital distance (Seager, 2013).

There is considerable uncertainty in the originally emplaced values of sulfur in the three terrestrial inner Solar System rocky planets, and some models suggest variations based on different source inventories or atmospheric stripping during planet formation (Javoy, 2010). These processes notwithstanding, the lack of long-term standing oceans over the planetary histories of Venus and Mars may reasonably be suspected to emplace sulfur differentially in the accessible reservoirs of those planets, while the lack of Earth-like volcanism limits the steady state concentration of sulfur in Venus' and Mars' atmospheres (Fegley & Prinn, 1989), though Venus' atmosphere may have variable sulfur dynamics for other reasons (Vandaele *et al.*, 2017; Jiang *et al.*, 2024). According to this analysis, whatever sulfur was once near the surface of Mars, the planet long ago stopped inputting significant amounts of sulfur into the atmosphere, and whatever sulfur is near the surface of Venus is not returned efficiently to deeper reservoirs due to the lack of interaction with a hydrological cycle or plate tectonics (Johnson & Fegley, 2002). Venus then has a $10^7$ fold atmospheric steady-state excess sulfur concentration compared to Earth because sulfur is not efficiently returned to surface reservoirs due to the lack of solubilizing water (due to surface temperatures which do not permit the existence of water (Loftus *et al.*, 2019)), whereas Mars has a low steady-state sulfur concentration both due to low volcanic sulfur exhalation rates, rapid photodestruction, and rapid stripping rates due to Mars' low mass. Earth maintains its surface sulfur abundance due to balances between these mechanisms. Understanding the abiological mechanisms which add or remove sulfur to the atmospheres of terrestrial planets is worthy of scrutiny and may help unravel whether the sulfur species detected in extrasolar planetary atmospheres can be interpreted as biosignatures. For example, it has recently been claimed low volatile C-species abundances in exoplanet atmospheres may be markers of surface oceans or biospheres (Triaud *et al.*, 2023).

## 2. Materials and Methods

### *2.1 - The Global Sulfur Cycling Model*

The presented general model runs on code adapted from Laneuville *et al.* (2018) and Jusino-Maldonado *et al.* (2022), which enables the fluxes and quantities of $S_{red}$ and $S_{ox}$ to be tracked and quantified. Building from this past work, we present a box and arrow model, in which sulfur masses are moved using defined geochemical rules between geological reservoirs. Similarly to the approach to abiotic N cycling used by Laneuville *et al.* (2018), we divided the tracked sulfur content into two redox states which we defined as reduced sulfur ($S_{red}$, which includes $S_0$ and lower redox states of sulfur, dominated in mass by $S^{2-}$ species) and oxidized sulfur ($S_{ox}$, which includes all species above the $S^0$ redox state, dominated by sulfate, with oxidation state +6). By considering only the two major redox states of sulfur during its geochemical cycling on Earth-like planets, we can simplify the processes shaping the distribution of these significantly different chemical forms of sulfur whose cycling we aim to track.

Sulfur transfer between reservoirs corresponds to geochemical fluxes modeled on our current knowledge of terrestrial geochemistry (Laneuville *et al.*, 2018; Jusino-Maldonado *et al.*, 2022). The model's flow is shown in Figures 1 and 2, and indicates the distinct geochemical processes modeled to transfer $S_{red}$ and $S_{ox}$ between reservoirs - some of which also enable changes in sulfur redox state.

### *2.1.1 - Model Reservoirs*

The following reservoirs were modeled as discrete units in this model: the atmosphere (ATM), oceans (Oce), oceanic crust (OC), continental crust (CC), marine sediments (MS), and upper (UM) and lower mantle (LM). For each geophysical reservoir, we tracked the masses of $S_{red}$ and $S_{ox}$ over each model timestep.

### *2.1.2 - Initial sulfur reservoir masses*

A crucial starting consideration is the amount of sulfur on Earth that was provided by accretionary processes. Estimates for these values typically use chondritic sulfur values (e.g. 21,000 ppm S for ordinary chondrites, see Wedepohl (1984)) as a starting point, though it is suggested a good deal of sulfur was lost during Earth's violent accretion process (Li *et al.*, 2016). Typically, scaling Earth's mass to chondritic sulfur content values suggests a bulk content that requires a large sulfur component in the planet's core (Dasgupta *et al.*, 2009). Due to the large mass and density of Earth's core this inference introduces an uncertainty which is common for other elements including N and P (Laneuville *et al.*, 2018; Jusino-Maldonado *et al.*, 2022).

Modern bulk terrestrial sulfur content has been estimated using measurement of meteorites (Wedepohl, 1984), estimates of the core's density, and measurements of various more directly accessible reservoirs. Estimates for bulk Earth S content range from 0.56% (Dreibus & Palme, 1996) to 5.9% (Wedepohl, 1984), or $3.34 \times 10^{22}$ to $3.5 \times 10^{23}$ kg S, or over a factor of ten in uncertainty. A more refined estimate based on assumed ratios of meteoritic feedstocks provides an intermediate value of ~$1.19 \times 10^{23}$ kg S (Wedepohl, 1984).

Several authors have estimated the sulfur content of the core, giving a range from 1.7 - 15 weight% (Ahrens, 1979; Brown & McQueen, 1982; Kargel & Lewis, 1993). This corresponds to a sulfur mass in the core from $3.35 \times 10^{22}$ kg (Dreibus and Palme, 1996) to $8.96 \times 10^{23}$ kg (estimated by McQueen & Marsh (1966)) based on measured core density deficiencies), or over a factor of 20 in uncertainty. Using $2 \times 10^{24}$ kg as the mass of the modern Earth's core (Papuc & Davies, 2008) and $4.06 \times 10^{24}$ kg as the mass of the mantle

(assuming it makes up ~68% of Earth's mass (Lodders, 1998)), residual sulfur contents of various surface reservoirs can be modeled to estimate more dynamic surface reservoir sulfur contents.

Better constraints on the mantle and crustal components can be placed on more measurable surface sulfur reservoirs including mantle rocks, which average around 124 ppm (range 34-214 ppm, Dreibus & Palme, 1996), with slightly higher measurements offered by other authors using values for Mid-Ocean Ridge Basalts (MORB) mantle (*e.g.* 150 ppm (O'Neill, 1991) to 350 ppm (Sun, 1982)). Subtracting the minimum core estimate from the maximum total Earth S budget, and the maximum core estimate from the minimum Earth sulfur budget gives a deficit range of S masses, that we considered as a starting pool for the evolution of the sulfur cycle, of approximately $10^{19}$ to $3.17 \times 10^{23}$ kg S. Estimates for the whole mantle reservoir, which must dwarf all overlying reservoirs, range from $1.38 \times 10^{20}$ kg S (assuming 34 ppm S average, Dreibus & Palme, 1996) to $1.39 \times 10^{21}$ kg S (assuming 350 ppm, Sun, 1982).

Examination of the composition of the core and mantle have led to estimates of ~6350 ppm sulfur for the whole Earth, which assumes a large loss of volatile sulfur during accretion and suggests $3.79 \times 10^{22}$ kg S: 1.9 weight% S in the core (or $3.7 \times 10^{22}$ kg S, McDonough & Sun, 1995) and 0.03 weight % S in the mantle (or $1.02 \times 10^{21}$ kg S, Carlson, 2005). We use the value of 1.7% weight sulfur in the core from Dreibus and Palme (1996) in Table 2.

### 2.1.3 - ET input

Meteorites, comets and interplanetary dust particles (IDPs) have provided a flux of extraterrestrial (ET) material to the inner rocky planets which has been declining over time since the formation of the Solar System (~4.5 Ga) (*e.g.* Bottke & Norman, 2017) with this input flux approaching an asymptote (Kyte & Wasson, 1986). The ET input flux of sulfur was modeled over geological time using the exponential decrease dynamic shown in Equation 1, which is governed by an exponential decrease timescale over millions of years (see Chyba & Sagan, 1992; Laneuville *et al.,* 2018). ET input flux was divided among Oce and CC reservoirs according to their dynamic surface area.

$$\text{Equation 1:} \quad F(t) = F_0 + (F_1 - F_0) e^{-t/\tau}$$

Here, F(t) is the extraterrestrial sulfur input flux (kg S yr$^{-1}$) at time t (Ma after formation), $F_1$ and $F_0$ are the maximum (initial) and minimum (late-time asymptotic) fluxes (kg S yr$^{-1}$), and τ is the e-folding decay timescale (Ma) (Table 3).

### 2.1.4 - Plate Tectonics

Plate tectonics are thought to be important for life, the long-term stability of the lithosphere, hydrosphere and atmosphere (see for example Valentine & Moores, 1974) and thus are important for sulfur cycling, providing a link between shallow and deep Earth reservoirs. During subduction, sulfur-containing materials are mainly either returned to the mantle, accreted to the CC, or volatilized into the atmosphere. At the same time, the formation of new OC from UM and LM reservoirs transfers sulfur from mantle to surface reservoirs as Mid-Ocean Ridge Basalts (MORB) or Ocean Island Basalts (OIB), which forms new OC, or sulfur that becomes entrained in seawater. The formation of supercontinents, orogenic activity associated with plate motion, and building of mountain chains also all promote erosion activity and sediment production, and these processes change weathering intensity (release of $S_{ox}$ by dissolution of $S_{red}$) and impact global terrestrial climatic conditions (Maffre *et al.*, 2018).

While the timing of the initiation of modern plate tectonics remains uncertain (Palin *et al.*, 2020), recent studies have proposed different lines of evidence for its potential origins. Ning *et al.* (2022) interpret eclogite-facies mineral assemblages in ~2.9 Ga rocks as evidence for modern-type ocean–continent subduction by that time. Chowdhury *et al.* (2023) argue that continental crust–forming processes operated from ~4.2 to at least ~3.7 Ga, and they suggest this interval may represent a transition in tectonic regimes.. While there is considerable uncertainty surrounding the onset time of plate tectonics, we included tectonic cycling using this model starting at the formation of the Earth (~4.5 Ga). This parameter can be easily modified in this model but was not explored here; the present model generally results in stabilization of sulfur reservoirs in the first 0.5-1.5 Ga of Earth's history.

This "box and arrow" model only keeps track of the sulfur masses and fluxes between reservoirs, without taking into consideration other geophysical processes occurring in these reservoirs. However, for the CC, in order to estimate erosion fluxes of CC sulfur, it is important to model and track the emergent CC land area, subject to subaerial erosion processes (see section 2.1.7). For this, although poorly constrained for the early Earth (Korenaga, 2018), we parameterized the growth of subaerial CC area based on Armstrong (1981), as a two piece function, starting at an arbitrarily low non-zero CC area that grows linearly with time until it reaches ⅔ of the present-day (~40 % of the total global surface area) at 1.5 Ga after Earth's formation. For the following 3 Ga of this simulation, subaerial CC area grows at a slower pace, reaching the present-day ~40% emergent CC area at a simulation time of 4.5 Ga. This is the same CC growth model used in Jusino-Maldonado *et al.,* (2022), which was compared to a significantly distinct, exponentially-decaying, growth rate model, with both models yielding a similar steady-state P distribution across reservoirs, suggesting these global cycling models are independent of CC growth models.

### 2.1.5 - Subduction

In this basic model, the surface-to-interior recycling is parameterized as an effective 'subduction' flux. In a mobile-lid (plate-tectonic) regime, this corresponds to plate subduction (Stern, 2002), whereas in non-plate regimes, analogous downward transport (e.g., delamination/foundering or episodic overturn) may occur but is not explicitly represented here (Crameri *et al.,* 2016).

In this subduction flux , some of the plate-entrained $S_{red}$ and/or $S_{ox}$ is delivered to deeper reservoirs following a typical OC turnover timescale of 100 Ma as considered previously (Jusino-Maldonado *et al.,* 2022). Continental accretion occurs as an ancillary outcome of subduction due to tensional stress at convergent plate boundaries (Zheng &

Chen, 2016). Subducting OC and MS plate material may become emplaced to the overlying CC, carrying their sulfur content with them (and MS material may contain a different sulfur redox content ratio compared to underlying OC). Our model simulates this complex process of subduction with 2 consecutive steps: 1) The taking the fluxes of S entering the subduction from OC and MS and 2) distributing this S mass through a 3-way partition into three processes, (Accretion into CC, Arc Volcanism into CC and Outgas, and proper subduction down to UM) described below. The ratio of material entering the subduction process that ends up excreted as arc volcanism is estimated to be between 1/4 to 1/7 of the total subducted material (Wallace, 2005) and is set at ⅕ in our standard model configuration.

We considered in our basic model a fixed timescale for subducting materials (constant D in Table 3), corresponding to the present typical OC recycling timescale (100 Ma, see Müller *et al.,* 2008). We also explored varied recycling timescales of 50 and 150 Ma (see Figure SI1). We modeled the subduction of this surficial material as a three-way partition of subducting material:

1) A fixed fraction (given by the accretion efficiency, **ε**, parameter) of subducted sulfur in MS is accreted onto the CC (with no chemical alteration)

2) Of the remaining MS and subducting OC, a fraction (**φ**, Table 3) of this subducting material is incorporated into the CC via arc volcanism. In this model we considered that a fraction of the sulfur content processed in arc volcanism lava is oxidized and injected to the atmosphere as $S_{ox}$ (dominated by $SO_2$) - and the remaining sulfur content is incorporated into the CC as $S_{red}$ (Wallace*,* 2001; Zelenski *et al.,* 2022).

3) The remaining MS and OC subducting material is subducted into the UM, and undergoes redox transformation in which all $S_{ox}$ is reduced to $S_{red}$, joining the reduced $S_{red}$ already present in the UM and/or LM.

### 2.1.6 - Atmospheric Rainout

The sulfur-containing volatile material outgassed to the atmosphere through different types of volcanism is generally rapidly removed from the atmosphere by photochemical reactions with other atmospheric components (Barnes *et al.,* 1986), most notably OH radicals. We consider here that the rainout of this atmospheric sulfur occurs uniformly across the planet's surface, being allocated to the CC or Oce reservoirs according to their relative surface areas, which change over time in this model. Although in modern Earth's atmosphere, carbonyl sulfide (OCS) is the most abundant S-bearing gas species, it is thought to primarily originate from dissolved marine organic matter of biological origin (Watts, 2000) and anthropogenic activities (Lee *et al.,* 2016). Chin & Davis (1993) estimate a total abiotic outgas of OCS (due to volcanism) of 3 x 10$^8$ mol/year, 3 orders of magnitude below $SO_2$ volcanic outgas rates, with OCS possessing an atmospheric lifetime up to 1 order of magnitude greater than $SO_2$. Hence, in our model we consider $SO_2$ as the main S-bearing atmospheric compound in a lifeless Earth atmosphere.

Unlike all other geochemical fluxes considered in this model, atmospheric rainout of sulfur compounds follows characteristic timescales of days, rather than thousands or millions of years, as the atmospheric half-life of $SO_2$ in the present troposphere and stratosphere of Earth is less than three months (Zhu *et al.,* 2020), and for $H_2S$, the present atmospheric half-life is only a few hours (Slatt *et al.,* 1978). Since the main atmospheric species limiting the lifetimes of $SO_2$ and $H_2S$ are OH radicals (Kasting *et al.,* 1989), whose abundance depends mostly on the humidity of the atmosphere and thus the presence of surface water and the atmospheric temperature profile, we did not parameterize these species loss rates as a function of the redox state of the atmosphere, but rather as simply "rapid removal" over the same timescales.

To model these significantly shorter timescales for the atmospheric lifetimes of sulfur compounds, we constructed a second temporal cycle (within each general simulation 5000 year timestep) with short timesteps equal to the simulation geochemical parameter of the atmospheric half-life of $SO_2$ ($\tau[SO_2]$) in the model's atmosphere. We henceforth call these "atmospheric timesteps." Assuming volcanic outgassing to the atmosphere is constant over each 5000 year timestep, at each atmospheric timestep *t,* the model inputs to the atmosphere the fraction of volcanic $SO_2$ corresponding to the time interval of the atmospheric timestep. As described in Equation 2, to the sum of the previous atmospheric timestep's atmospheric mass of $SO_2$ with the volcanic input over one atmospheric timestep, half of this sum is removed through rainout per atmospheric time step, and added to the CC and Oce reservoirs, proportional to their relative global surface coverage. Each atmospheric time step thus corresponds to one $SO_2$ atmospheric half-life.

$$\text{Equation 2: } [ATM\ SOx](t+1) = \frac{[ATM\ SOx](t) + Volcanic\ Input\ of\ SOx\ (t+1)}{2}$$

Here, $[ATM\ SO_x](t)$ is the atmospheric mass of oxidized sulfur (kg S) at atmospheric timestep t; Volcanic Input of $SO_x$ (t+1) is the volcanic oxidized-sulfur input in the t+1 atmosphere timestep (kg S), and the factor of 1/2 represents removal of half the atmospheric burden per timestep (one $SO_2$ half-life, $\tau[SO_2]$, in years; Table 3). The removed $SO_x$ (that does not remain in the atmosphere) undergoes rainout, being allocated to Continental crust and oceans proportionally to surface area.

S flux to the atmosphere via wind erosion and sea spray presently contributes small amounts of sulfur to the global flux (*e.g.* aeolian erosion is estimated to presently mainly contribute anthropogenic S, which has a short atmospheric lifetime due to rapid removal rates (Bates *et al.,* 1992), while sea spray has been estimated to contribute 1.3 - 2.8 x $10^{11}$ kg S $y^{-1}$ (Varhelyi & Gravenhorst, 1983) which is mainly trapped in the marine layer, and has a short atmospheric lifetime (Bates *et al.,* 1992)), not relevant at the timescales of thousands or millions of years discussed in this study. We thus did not include these fluxes in our model.

### *2.1.7 - Riverine Erosion and Sedimentation*

The three most dynamic reservoirs in the geochemical sulfur cycle (CC, Oce, and MS) are highly connected via erosion and sedimentation. Erosion is assumed to be dominated by riverine and other hydrologic processes, and comprises the main sulfur removal process from the CC in this model. Erosion allows the transport of sulfur-bearing CC into the Oce reservoir, which can then transport to the MS reservoir. We modeled the erosion fluxes of $S_{red}$ and $S_{ox}$ from CC (in kg $yr^{-1}$) as a product of emergent land area (A) and S redox species concentration ($S_{red}$ or $S_{ox}$) by a measure of average global erosion rate (**α**) with m/year units as:

Equation 3: $Flux_{Eroded} = \alpha \cdot A \cdot \rho_{CC} \cdot [S_{red/ox}]$

Here, $Flux_{Eroded}$ is the eroded sulfur flux (kg S yr$^{-1}$), α is the mean physical erosion rate (m yr$^{-1}$; Table 3), A is emergent continental land area (m$^2$), $\rho_{CC}$ is mean continental crust density (kg m$^{-3}$), and [$S_{red/ox}$] is the mass fraction of reduced S or oxidised S in continental crust (kg S per kg crust).

with $\rho_{CC}$ as the mean density of the CC. This process includes a chemical redox transformation rather than merely a simple flux transfer of sulfur between reservoirs. Consequently, on an oxidized post-GOE Earth-like planet (see below, section 2.2), a significant fraction of CC-eroded $S_{red}$ is oxidized during environmental transport by the time it reaches the ocean as Oxidatively Weathered Sulfide (OWS), see Nordstrom (2011). Presently, between CC source rock erosion and hydrological transport to the oceans, the vast majority of CC $S_{red}$ is oxidized by physical and biological agents (Calmels et al., 2007, Burke et al., 2018), leading to delivery of only a miniscule amount of eroded CC $S_{red}$ to the oceans, mainly in the form of sedimentary sulfide (Hofmann et al., 2009; Reinhard et al., 2013; Da Costa et al., 2017). In the pre-GOE-conditions model (which we refer to as "no-OWS scenario", see section 2.2), we assume that no OWS is removed from the CC, meaning that only eroded CC $S_{ox}$ and non-OWS CC $S_{red}$ (e.g. the amount of $S_{red}$ that is soluble in equilibrium with mineral phases) is delivered to the Oce reservoir.

As this CC erosion sulfur flux enters the Oce reservoir, it becomes dissolved Oce $S_{red}$ or $S_{ox}$ - until it exceeds the Oce carrying capacity for $S_{red}$ or $S_{ox}$ as determined by model-permitted solubilities estimated from measured $K_{sp}$ values and commonly encountered marine conditions of pH, salinity, alkaline earth metal concentrations, temperature and pressure, bearing in mind that these simplify complex conditions in which kinetic or thermodynamic considerations may govern solubility (e.g. Kopittke et al., 2004). A more complete analysis of this question would consider $S_{red}$ or $S_{ox}$ species solubility as a function of the concentrations of counterions (such as transition metals and alkaline earth metals), which are themselves dependent on environmental conditions which control complexation with other counterions that govern S-species solubility, and which are themselves subject to other systemic feedbacks. Admittedly, especially neglected in this analysis is the role of iron for which sulfur complex solubility is tightly linked to iron redox speciation, which is likewise linked to the redox environment.

As $S_{red}$ or $S_{ox}$ concentrations exceed the saturation limits of the Oce reservoir in the model via input from other reservoirs, excess $S_{red}$ or $S_{ox}$ is precipitated from the Oce reservoir to the MS reservoir. A $BaSO_4$ (barite) $S_{ox}$ removal mechanism was included in this model since barite deposits are known dating back to the Archean (Reimer, 1980), when Oce sulfate concentrations were by most estimates very low (Crowe et al., 2014). The reaction of $S_{ox}$ with supply-limited Ba was used as a control of this flux since barite is especially insoluble ($K_{sp}$ ~ 10$^{-8.2}$ at 25° C, (Rushdi et al., 2000; Jewell, 2000); contrasted with a value of ~ 2.6 x 10$^{-5}$ for gypsum ($CaSO_4 \cdot 2H_2O$)) and thus provides a constant background precipitation removal mechanism for low Oce [$S_{ox}$]. While the greatest mass of $S_{ox}$ sedimentary deposits in the geological record is in the form of gypsum, which appears to have precipitated in high [$S_{ox}$] post-GOE evaporative marine environments, barite deposits are common in modern marine sediments (Johnson et al., 2017) and extend back billions of years (Reimer, 1980).

The geochemistry of barite is complex and several types of barite formation mechanisms are documented, nevertheless the majority, at least in recent environments, appear to be formed by precipitation in the marine water column (Johnson *et al.*, 2017). Comparison of the estimated riverine $Ba^{2+}$ delivery to the modern oceans (~2 x $10^8$ kg S equivalents $y^{-1}$ (Henderson & Henderson, 2009)) roughly matches the estimated marine barite sediment accumulation rate within one or two orders of magnitude (Paytan *et al.*, 1996), which is a good agreement given uncertainties in changes of continental surface area, erosion rates and ocean chemistry. Inclusion of barite precipitation provides a mechanism for the removal of low concentrations of Oce $S_{ox}$ to MS reservoirs over geological timescales even when oceanic sulfate concentrations are below concentrations required for gypsum precipitation. Within a timestep of 5000 years, we expect that such a mechanism would remove a mass of baryte from the global oceans into MS equivalent to a total $S_{ox}$ mass of $10^{12}$ kg. This corresponds roughly to the removal of $10^{-6}$ of the total oceanic mass of $S_{ox}$ every 5000 years from the global ocean. Thus, in our standard model configuration, barite removal was parameterized to remove $10^{-6}$ of the total Oce $S_{ox}$ mass per timestep., Since at each timestep, the input of oceanic Sox mass depends on surface erosion, baryte removal is thus (indirectly) dependent on the surface erosion flux.

### *2.1.8 - Volcanism and Hydrothermal Activity*

Volcanism was modeled such that the S sources were the UM and LM reservoirs that distribute materials to sinks (CC, OC, Oce, and atmosphere) via three conduits: MOR, arc and hotspot volcanism. Presently, the estimated S flux added to the CC (which includes sulfur input from Atm, and OC and Oce input via tectonic accretion) through arc volcanism is ~0.5 to 5 $km^3$ $yr^{-1}$ (Deligne & Sigurdsson, 2015), whereas ~20 $km^3$ $yr^{-1}$ is added via MOR volcanism to OC and oceans and hotspot volcanism (which mainly derive from the LM) add ~2 - 2.5 $km^3$ $yr^{-1}$ to surface reservoirs (CC and atmosphere or OC and oceans), according to the ratios of CC to OC surface area. In this model, all three types of volcanism add $S_{red}$ to the OC and CC, whereas $S_{ox}$ is added to the Oce and atmosphere as volatile materials (see model flowchart in Figure 1). We added a sulfur enrichment factor by observing the variation in S-content between the mantle and various magmas (Wallace *et al.*, 2015, Table 3). These enrichment factors were included to reflect how sulfur concentrates in various magmatic environments. Also following Laneuville *et al.* (2018) and Jusino-Maldonado *et al.* (2022), based on Turcotte (1980), we considered that the long-term thermal evolution of Earth has caused a decrease in volcanic activity by volume of erupted material since its formation. We parameterize this effect using an exponential decay as described by Eq.1, with a typical exponential decay timescale of 150 Myr from an erupted volume three times higher than at present (see Table 3).

This model considers the difference in sulfur dynamics and redox state between sulfur outgassed by subaerial and submarine volcanism. Modeled fluxes track the differential transfer of sulfur species from the BSE to the oceans and atmosphere, allowing consideration of the effects of volcanic outgassing in submarine environments, where presently ~0.7–1.4 × $10^{11}$ mol $yr^{-1}$ S at submarine arc volcanoes (Butterfield *et al.*, 2011) and ~2.67 x $10^{12}$ mol $yr^{-1}$ S at MOR (Fischer, 2008) are outgassed into the oceans. Butterfield *et al.* (2011) noted that contemporary submarine arc volcanoes produce solutions that are ~100 mM in sulfite (derived from $SO_2$) and 30 µM in dissolved sulfide, and that a large percentage of the elemental and reduced sulfur discharge is immediately deposited in the OC and MS in the vicinity of such submarine volcanoes.

In our basic model, a fixed 20% of the submarine volcanic sulfur output in hotspot and MORB volcanism is incorporated into the Oce reservoir as dissolved $S_{ox}$ or $S_{red}$ (Oppenheimer *et al.,* 2011). Modeled subaerial eruptions of hotspot and arc volcanism applied a fraction of mantle sulfur as input to atmospheric $S_{ox}$ (fixed at 50% of erupted sulfur mass in the case of arc volcanism, following Wallace, 2005) to account for the present-day ~$3.15 \times 10^{11}$ mol yr$^{-1}$ of sulfur volatiles input into the atmosphere by subaerial volcanism (Fischer 2008).

Despite the wide array of sulfur-bearing species at distinct redox states known to be outgassed in volcanic settings (*e.g.* $SO_2$, OCS, $H_2S$, *etc.* (Oppenheimer *et al.,* 2011)), the comparatively short atmospheric half-lives and smaller abiological production rates of reduced sulfur volatile species compared with $SO_2$ on modern Earth led us to treat all volcanic outgassing as occurring in the form of $S_{ox}$. Even in pre-GOE scenarios, calculations suggest that $SO_2$ was marginally more stable in the mildly-reducing atmosphere postulated for the archaean Earth than $H_2S$ (Kasting *et al.,* 1989).

Submarine hydrothermal activity is another process included in this sulfur cycling model. Albeit smaller in flux to volcanism or subduction, sulfur compounds are commonly found in deep marine environments in various oxidation states (Alain *et al.,* 2022). Presently the entire ocean volume is estimated to circulate through marine hydrothermal systems approximately every $10^7$ years (Holland, 1984). Modern hydrothermal circulation efficiently removes dissolved oceanic sulfate (Canfield, 2004; Alt *et al.,* 1989), causing the precipitation of inorganic sulfate into the OC, largely as anhydrite. Similarly to volcanism, the efficiency of this process is assumed to have exponentially decayed over geological timescales (see Table 3). Entrained seawater sulfate is abundantly deposited as anhydrite along hydrothermal vent axes, due to anhydrite's retrograde thermal solubility (Alt, 1995; Blounot & Dickson, 1969). Indeed anhydrite appears to be enriched in hydrothermal zones where cool sulfate-bearing seawater mixes with ascending hotter basement fluids, such as the upper part of sheeted dyke complexes (Alt *et al.*, 2010). According to Alt (1995), most of the anhydrite that forms in high-temperature axial seawater circulation is likely later dissolved during lower temperatures off-axis circulation, with the sulfate being returned to seawater and thus reducing the net uptake of S in oceanic crust by some unknown amount. Evidence for this includes the presence of anhydrite pseudomorphs after partial dissolution of anhydrite crystals (Alt *et al.*, 1989; Davis *et al.*, 1992). Such open circulation of seawater through the upper oceanic crust may persist on average for 65 Ma (Stein & Stein, 1994), which is considerably shorter than the mean estimated modern lifetime of OC.

Since it is not clear how much of the anhydrite precipitated during on-axis hydrothermal circulation of seawater sulfate ultimately makes it to subduction zones (without undergoing resolubilization as the oceanic crust cools), we consider that the hydrothermal circulation rate can be considered as a proxy for the dissolution rate of precipitated anhydrite, *e.g.*, if the mean ocean hydrothermal circulation time is one million years and no anhydrite redissolves this is equivalent to the oceanic crustal emplacement of anhydrite when the mean circulation time is ten million years and 90% of emplaced $S_{ox}$ redissolves and so on.

We have thus modeled oceanic $S_{ox}$ removal into OC through hydrothermal circulation per timestep by removing into OC the fraction of oceanic $S_{ox}$ mass that has undergone hydrothermal circulation in that timestep. This corresponds to a fraction of oceanic $S_{ox}$ mass being removed equal to the inverse of the timescale required for the entire ocean to circulate through marine hydrothermal systems. Resolubilization is taken into account by increasing the mean circulation time in the model by a factor corresponding to the ratio between the originally precipitated mass of $S_{ox}$ through hydrothermal circulation and the mass of precipitated $S_{ox}$ into the oceanic crust that is not resolubilized before undergoing subduction. Given the considerable uncertainty regarding this value, we assumed it corresponds to a net decrease in efficiency of hydrothermal circulation removal by a factor of an order of magnitude - decreasing the effective timescale for hydrothermal circulation to $10^{-8}$ $yr^{-1}$. At the parameter sweeping section of this work (section 3.2), we explore other possible values for this parameter, and assess the model's sensitivity to it.

In our basic model, sulfur introduced via HT vent weathering of mantle-derived rocks is rapidly deposited in submarine environments and added to the MS reservoir (Alain *et al.*, 2022), while anhydrite (represented as $S_{ox}$) precipitates via HT circulation and is added to the OC reservoir. Functionally this decision has no impact on the model, since OC and MS are equally disproportionated during subduction to outgoing reservoirs. Such a distinction might be meaningful for a model that considers biological sulfur redox activity. This model likewise combines $S_{red}$ and $S_{ox}$ in MS and OC for the sake of their release during arc volcanism, as these two pools are combined, and likely redox equilibrated with other species in subducting melts.

### *2.1.9 - Core, lower mantle and upper mantle*

The primary process controlling the sulfur distribution in smaller reservoirs in this model is mantle convection (due to the large mass of the mantle), which remains poorly understood. Mantle convection is believed to play a pivotal role in determining plate velocities, sea-floor subsidence, volcanism, gravity anomalies, and other geophysical phenomena (Bercovici & Mulyukova, 2021). In this model, a two-layer mantle is assumed, in which mantle convection facilitates the mixing of sulfur from the upper mantle with the deeper mantle (with sulfur contents of both layers acquired via subduction or surface reservoirs). While the actual stratification of the mantle remains debatable, and it remains unknown how mantle stratification may have evolved over time, we used a simplified model assuming a two-layer mantle model which assumes some kinetic inertia between rapidly mixed and slowly mixed mantle components (Bercovici & Mulyukova, 2021).

Mantle convection parameters are difficult to constrain, but it should be noted that the mantle in general has a great deal of "model inertia" due to its size, and exchanges material relatively slowly to other reservoirs, thus it takes large parameterization changes in mantle dynamics to significantly affect the model output. We introduced a parameter in our simulation (mantle mixing rate, see Table 3), which idealizes the mixing of upper and lower mantle mixing within a certain range of timescales.

It is also debated whether and to what degree material is exchanged between the core and mantle (Brandon & Walker, 2005). This model assumes the core is physically "closed" from interacting with the mantle over the timescales considered in the model.

### 2.2 - Modeling the Impact of the GOE on Sulfur Cycling

There remains some debate of the timing and cause of the GOE, and whether it was biologically or geologically caused (Chen *et al.*, 2022). While we remain agnostic to this question, this model allows exploration of scenarios in which a GOE-like surface oxidation occurs via purely abiological mechanisms (e.g., in which the generation of surface oxidants by abiological means overwhelms the ability of tectonic processes to titrate them with reductants), which may be useful for future investigations (Wogan *et al.*, 2022).

We additionally explored the impact of the generally accepted biologically-mediated transition between the pre-GOE and post-GOE Earth (Kharecha *et al.*, 2005; Lyons *et al.*, 2014). We constructed two distinct model configurations exploring two end-members of possible Earth-like terrestrial planet geochemistry: one mimicking pre-GOE abiotic sulfur cycling, and another mimicking post-GOE abiotic sulfur cycling (still without considering any contributions of biological processes, considering an *ad hoc* oxidized planetary state on a lifeless Earth-like terrestrial planet, akin to a lifeless, post-GOE Earth, *e.g.* assuming the GOE was the result of Earths' internal abiotic redox dynamics titrating its surface chemical dynamics). These models allow for independent parameterization of chemodynamics for each distinct planetary redox state.

Following Reinhard *et al.* (2009), Stueken *et al.* (2012) and Wang *et al.* (2019) we considered that the most important changes in abiotic geochemical sulfur cycling that took place during a GOE-like transition occur in the CC → Oce pathway, as drastically distinct environmental redox states strongly impact the erosion fluxes of different sulfur redox states. Hence, the difference between pre and post-GOE-like model conditions corresponds to the inclusion (or exclusion) of Oxidatively Weathered of Sulfide (OWS, Reinhard *et al.*, 2009) in the model. For the scenario mimicking post-GOE conditions, we considered that a significant fraction of erosion was spurred by CC OWS. This fraction was set to 99.99%, meaning that out of the eroded material, on the standard model configurations, only 0.01% of eroded CC $S_{red}$ would reach the oceans as $S_{red}$, with the remaining being oxidized into $S_{ox}$ (as defined by the parameter "Fraction of non-oxidized Sulfide during Surface erosion" in Table 3). Given the uncertainty associated with this parameter, we performed a parameter sensitivity analysis on the model with respect to this variable in section 3.2 (see Figure SI3). Adding OWS to the model causes a significant increase in the CC output of erosion flux as $S_{ox}$, which is added to the previously considered erosion of CC $S_{ox}$ and erosion of CC $S_{red}$ (Reinhard *et al.*, 2009; Wang *et al.*, 2019). We henceforth call this the "high-OWS scenario".

For a scenario mimicking pre-GOE conditions, we considered that no chemical alteration in the redox state of eroded materials occurs, and hence, all eroded CC $S_{red}$ reaches the ocean as $S_{red}$. Thus in this case, the fraction of $S_{red}$ reaching the ocean during surface erosion is set to 100%. We henceforth call this the "no-OWS scenario". Since under post-GOE conditions, the bulk of S erosion is OWS-driven, removing OWS in a pre-GOE scenario will necessarily affect the net erosion rates. Hence to make our model self-consistent in the pre-GOE model scenario, we scale up the "fraction of surviving Sred reaching the ocean" from 0.1% to 100% while scaling down the net erosion rates by the same factor. This implies that in a pre-GOE scenario, the net flux of $S_{red}$ from CC into the ocean is comparable to that of the post-GOE Earth, but all OWS flux into the oceans stops.

We did not consider in detail how a GOE would alter the oceanic solubility of sulfur compounds (for example by affecting the solubility and transport dynamics of transition metals such as iron, whose solubility would affect $S_{red}$ species solubility) or the atmospheric half-lives of sulfur volatiles (*e.g.* via shielding by $O_2$-derived ozone), though it is clear that the oxidation state of planetary atmospheres impacts the lifetime of atmospheric sulfur species considerably (Kurzweil *et al.*, 2013). Despite this, the photochemical models of Kurzweil *et al.*, (2013) point to atmospheric concentrations of sulfur-bearing compounds that do not vary significantly as $pO_2$ increases by six orders of magnitude.

The solubility of iron in anoxic seawater is also photochemically governed (Konhauser *et al.,* 2007), and the solubility of sulfate minerals is largely pH- and alkaline earth metal-solubility governed (Freyer & Voight, 2004), which thus connects sulfur to the carbon cycle, since $Ca^{2+}$ concentrations are likely governed by other geochemical processes (Lebrato *et al.,* 2020). Despite the large uncertainties, it has been shown (through the comparison of dissolution and sedimentation timescales of sulfide minerals in the Archean oceans with variable potential $O_2$ concentrations) that sedimentation prevails over oceanic dissolution of sulfide minerals (Reinhard *et al.,* 2013). This shows that rather than chemical alteration and dissolution, the oceanic component of the reduced sulfur cycle is dominated by sedimentation of detrital pyrite in a pre-GOE ocean, and by the sedimentation of sulfate originating from abundant continental OWS. Hence, the major effect of a GOE, whatever its source, on a hydrologically active planet's sulfur cycle is likely the mobilization of subaerial mineral sulfur via OWS. The amount of oxygen in Earth's present atmosphere, mainly derived from oxygenic photosynthesis, (~3.7 x $10^{19}$ moles $O_2$, sufficient to oxidize ~1.85 x $10^{19}$ moles or ~5.9 x $10^{17}$ kg S) pales in comparison with the sizes of terrestrial sulfur reservoirs (see Table 3), again pointing to sulfur cycling being a dynamic means of shuttling electrons to the much larger planetary iron reservoir.

### *2.3 - Random seeding vs Parameter Sweeping*

To test our model for its ability to produce novel output phase spaces based on random initial conditions, we conducted two simulations following previous approaches (Laneuville *et al.,* 2018; Jusino-Maldonado *et al.,* 2022).

First, we were primarily interested in probing how dynamical box models evolve over geological time for an ensemble of distinct initial conditions, for fixed and dynamic geochemical parameter configurations. Hence, a set of randomly seeded initial conditions was included for the initial $S_{ox}$ and $S_{red}$ mass for each reservoir, allowing each of the reservoirs' initial masses to be set between zero kg and the modern estimated BSE mass, such that the total system sulfur mass was equal to the modern estimated BSE mass, accounting for uncertainties in Earth's initial sulfur endowment and early differentiation. We fixed the initial mass of CC and MS as zero, conditions that while unlikely, allow these reservoirs to grow according to system parameterizations over the course of the 4.5 Ga simulation. Our end goal with random seeding simulations was to examine whether after 4.5 Ga of simulation time the reservoirs $S_{ox}$ and $S_{red}$ masses would converge to a steady state or develop divergent trends, rather than whether they would converge on modern values. We additionally hoped to compare the final sulfur mass distributions of an abiological sulfur cycling model with present-day Earth reservoirs' sulfur content, as a way to detect possible impacts of life on the planetary distribution of distinct sulfur redox states in detectable reservoirs.

Second, we aimed to explore how varying key parameters governing the modeled geochemical processes would impact the final reservoir mass distribution. For this, we ran simulations with constant initial mass distribution, varying one geochemical parameter at a time. Observing how the reservoir mass distribution after 4.5 Gyr was affected by varying a geochemical parameter presents an indication of its impact on the overall sulfur cycle - while also allowing constraint of the impacts of geological parameter uncertainty on the final simulation outcomes. Plots showcasing the effect of geochemical parameter variation on final reservoir mass distribution are shown in images SI1 to SI6.

Note that the summed values in Table 2 estimate a slightly higher total S mass in terrestrial reservoirs than is estimated using chondritic accretionary values provided in Table 1 (3.61 x $10^{22}$ vs. 3.34 x $10^{22}$ kg, respectively, or a difference of ~8%; summing the total reservoir masses also overestimates the total planet mass by ~ 3%). While this is likely due to rounding errors, it is likely that whatever the initial inventory of sulfur initially accreted from ET materials, some sulfur was lost by impacts and atmospheric erosion. This underscores the uncertainty of estimating initial sulfur inventories, but does not significantly impact this model since the vast majority of the starting sulfur in the model is either emplaced in the core, which does not functionally engage in the model, or emplaced in the mantle, which has enormous kinetic inertia in the model due to its large size and slow turnover.

## 3 - Results and Discussion

### 3.1 - Random Seeding and Temporal Evolution Models

Following previous work (Laneuville *et al.*, 2018; Jusino-Maldonado *et al.*, 2022), we modeled the mass flux of two idealized sulfur redox states ($S_{red}$ and $S_{ox}$) across distinct planetary reservoirs over the course of 4.5 Gyr. Upon selecting a set of best-guess ensemble of geochemical parameter values (detailed in Table 3), we present here the results of the model's temporal evolution from wide varied initial conditions of reservoir masses, accounting for the significant uncertainties regarding the original sulfur mass distribution among terrestrial reservoirs after the Moon-forming impact. The temporal evolution of these varied initial seedings are shown in Figure 3 for a set of geochemical parameters corresponding to the no-OWS scenario, which mimics an abiotic pre-GOE Earth-like planet, and Figure 4 for a set of geochemical parameters corresponding to the high-OWS scenario, which mimics an abiotic post-GOE Earth-like planet.

Our models suggest that even on planets on which life fails to arise, a wide range of initial conditions converge on sulfur distributions resembling the modern terrestrial sulfur cycling regime after a characteristic timescale of about 500 Ma (see Figures 3 and 4) up to

2.5 Ga for the CC, suggesting biology has had generally subtle, but in places dramatic, effects on the long term evolution of the sulfur cycle. Beyond these timescales, most reservoir masses converge on abiological steady states. An exception is found in the steady increase in atmospheric $S_{ox}$ mass in simulations in both no-OWS and high-OWS conditions (see Figure 4f). For no-OWS model simulations, the $S_{red}$ contents of MS and CC and atmospheric $S_{ox}$ mass do not reach a steady state after 1.5 Ga of simulation (Figures 3c, 3d and 3f) continuing their mass growth (by less than an order of magnitude) for the last 3Ga of simulation. This seems to be counterbalanced by slight mass decreases on deeper BSE reservoirs (OC, UM and LM) during the last 3Ga of simulation . The main sink of CC sulfur in our model is OWS and its shutdown (in a no-OWS model), may contribute to a build up of CC $S_{red}$ mass, which subsequently leads to the $S_{red}$ build up in the MS reservoir over time.

Comparing the final reservoir masses of $S_{ox}$ and $S_{red}$ after 4.5 Ga of simulation with current estimates for present-day Earth's reservoir masses of sulfur species (Table 2), it is possible to postulate the impact of Earth's biosphere on global sulfur cycling. Mantle reservoirs, due to their large masses, stabilize at reservoir mass values with relatively reduced deviations from present Earth levels (PEL), to less than 20% deviance from PEL (Figures 3a and 4a). This is a small deviation considering the uncertainties surrounding current estimates for the total sulfur content of Earth's mantle (Carlson, 2005), with some models varying by an order of magnitude. In reservoirs with lower sulfur mass, such as the OC or CC, these deviations between estimations of PEL and simulations can be greater in relative terms, (see Figures 3b and 4b for OC and 3c and 4c for CC) but still within the same order of magnitude.

Compared to what was shown in Jusino-Maldonado *et al.* (2022) regarding simulated P cycling using a similar model, in this model, the Atmosphere and Oce reservoirs behave as small and transient reservoirs, which significantly impact the global transfer of sulfur by enabling rapid interconnection between reservoirs. Given the large size of the fluxes interacting with these reservoirs relative to their transient mass, we expected significant deviations from their PEL levels. This is the case for the atmospheric reservoir mass of $S_{ox}$, (see Figures 3f and 4f) which reaches the end of the simulation at a mass value smaller than present-Earth estimates by a factor of two. This is not the case for the Oce reservoir, as independently from the initial reservoir mass distribution, this reservoir's mass quickly (in less than 100 Myr of simulation time) stabilizes at a steady-state mass with a deviation from present-day ocean mass values of $S_{ox}$ and $S_{red}$ below 10% (see Figures 3e and 4e). This suggests that the ocean reservoir's S fluxes are dominated by the erosional input of CC S tightly coupled with precipitation onto marine sediments, such that the oceans remain close to a saturated state for both $S_{ox}$ and $S_{red}$ species akin to modern oceans (Meijer & Van Rosmalen, 1984; Al-Farawati & van den Berg, 1997). These results may have important implications for icy moons where the oceanic reservoir is not a small or transient reservoir, but rather a steady-state component of these world's sulfur-trading reservoirs, and in which many of the reservoir fluxes described here would be inoperative (Coelho & Martins, 2021; Xu *et al.*, 2025).

The most striking deviation from PEL reservoir steady state masses in our simulations is observed for the MS reservoir (Figures 3d and 4d). Unlike the present Earth, our model stabilizes with a $S_{ox}$ MS mass four orders of magnitude larger than its $S_{red}$ counterpart, with the model's $S_{red}$ MS mass being four orders of magnitude below the PEL level. This is a consequence of the dominant form of erosion of sulfur-bearing minerals of the mostly $S_{red}$-containing CC reservoir acted upon by Oxidatively Weathered Sulfide (OWS), as the heavily eroded and weathered sulfides on the high-OWS scenario are transported and deposited into the MS reservoir as $S_{ox}$ (Reinhard *et al.*, 2009; Wang *et al.*, 2019).

This result is also observed for the no-OWS simulations, which mimic pre-GOE conditions, in which OWS was turned off in the model. Taking into account the considerable uncertainties of pre-GOE erosion rates (Kambler *et al.*, 2001), we modeled the final reservoir masses for a wide array of erosion rates (see the parameter sweeping plots in Figures SI2g and SI2h). These show that even an extreme two order of magnitude increase in erosion rate fails to increase the $S_{red}$ MS mass above 10 % of PEL even for the model configurations without Oxidative Weathering of eroded CC Sulfide (no-OWS simulations, mimicking pre-GOE Earth). Interestingly, this effect may be explained by the impact of life on our model. The impact of microbial sulfate reduction (MSR) in sedimentary settings has been described in previous studies (Habicht & Canfield, 1997; Wasmund *et al.* 2017; Jorgenssen *et al.*, 2019), which have established the importance of the marine microbiome on the reduction of sulfur species in organic-bearing sediments. Our model supports the idea that life is responsible for the redox-state distribution of sulfur in sedimentary environments, and that in its absence, sedimentary deposits should contain a far more oxidized inventory of sulfur species, although these may not be stable over geological time scales.

### *3.2 - Parameter Sweeping and Sensitivity Analysis*

In regard to the model's sensitivity to the key geophysical and geochemical parameters considered, we tested the impact of varying erosion rates for distinct Oceanic Crust recycling timescales (Figures SI1), varying erosion rates for distinct OWS model configurations (no-OWS versus high-OWS, figures SI2) and varying erosion rates for distinct fractions of non-oxidized sulfide during OWS (figures SI3). We have also tested the impact of varying baryte precipitation fractions of oceanic sulfate for distinct Solubility Limits of Sulfate in the Ocean (figures SI4) and baryte precipitation fractions for distinct hydrothermal circulation rates at the present (figures SI5). Finally, we also tested the impact of varying volcanic outgassing efficiency for both no-OWS and high-OWS scenarios (figures SI6) . This sensitivity analysis study allows not only better comprehension of the model by exploring its parameter space, but it also provides an indication of the steady-state result's error bars associated with parameter uncertainty.

Similarly to what had been observed in Jusino-Maldonado *et al.* (2022) for an abiotic P cycling model, some steady state reservoir masses of sulfur species are highly sensitive to global erosion rates (figure SI1). This is the case for CC reservoirs of $S_{red}$ and $S_{ox}$, for which increasing erosion rates by an order of magnitude is enough to decrease steady state reservoir masses by at least one order of magnitude (figures SI1e and SI1f). This mobilized Sulfur content is transported to the MS reservoir - in particular, for MS $S_{ox}$ due to the effect of oxidative weathering of CC sulfide under high-OWS conditions. This enables MS $S_{ox}$ masses to increase by up to 50% (figure SI1h) as erosion rates increase by 3 orders of magnitude in the vicinity of PEL, showcasing a tight coupling between CC and MS mediated by riverine erosion. This effect is not observed on oceanic crust $S_{ox}$ masses, which does not showcase a dependence on erosion rates (see figure SI1c and SI1d), showcasing the relative dominance of other oceanic $S_{ox}$ sinks, such as precipitation into MS. Similarly, the steady state sulfur masses of the oceans are not significantly affected by erosion rates (see figures SI1i and SI1j), pointing to the strong effect solubility limits, baryte removal and hydrothermal circulation removal produce in the ocean reservoir - turning this reservoir into a pathway that accelerates sulfur transfer downward from continents to marine sediments. Through subduction, an increased mass of MS S-bearing species also lead to higher downward flow into the mantle reservoirs (see figure SI1a and SI1b), increasing UM S masses by up to 2% as erosion rates increase by 3 orders of magnitude.

OC recycling timescales (a proxy for subduction timescales, which are inversely proportional to subduction rates) have strong effects on OC and MS reservoirs: as more efficient subduction processes prevent larger mass build-ups in these reservoirs (see figures SI1c, SI1d, SI1g, SI1h). Conversely, higher mobility rates of surface material (linked to higher erosion rates and lower OC recycling timescales) increase the transport of surface material to mantle reservoirs, as suggested in Jusino-Maldonado *et al.* (2022) - leading to possible increases of UM $S_{red}$ mass which could not exceed more than 2% of PEL values (see figure SI1b).

Another phenomenon observable in the parameter sweeping plots (SI2) with the comparison of how reservoir sulfur mass depends on geochemical parameters for the no-OWS and high-OWS model configurations is the relative change in sulfur steady-state distribution before and after a global oxidation event akin to Earth's GOE. Figures SI2 show this comparison for $S_{ox}$ and $S_{red}$ steady state reservoir mass as a function of erosion rate for no-OWS and high-OWS model configurations. These plots highlight the transfer of sulfur from surface reservoirs to deeper reservoirs via GOE-mediated OWS, as deeper reservoirs (LM, UM, OC) become more sulfur-enriched in post-GOE-mimicking, high-OWS scenarios, while surface reservoirs such as CC show higher sulfur enrichment in no-OWS scenarios. This is a consequence of the increased mobility of crustal sulfur with the onset of OWS, dominating over the redeposition rates of sulfur in crustal reservoirs (via volcanism and continental accretion) showing the impact of OWS on the sulfur cycle (Reinhard *et al.,* 2009; Wang *et al.,* 2019). As for MS reservoirs, the onset of OWS leads to a decrease of $S_{red}$ MS mass of three orders of magnitude for present-day erosion rates, and even for erosion rates two orders of magnitude above present-day rates, $S_{red}$ MS mass does not reach above 10% of its present-day mass. This agrees with the fact that the geological record prior to the GOE provides evidence in the form of pyrite ($FeS_2$) grains in deltaic environments, which under the action of OWS, would have been oxidized and dissolved into the Oce (Schopf, 2014). With the onset of OWS, these eroded grains would swiftly be oxidized and deposited into marine sediments as $S_{ox}$ in greater quantities, given the more effective erosion of S-bearing minerals enabled by OWS.

This idea is further supported by another set of parameter sweeping plots (SI3), which assess the effect of OWS efficiency during high-OWS conditions on the steady-state global distribution of sulfur redox states, as a function of erosion rates. The model sensitivity analysis for this parameter allows us to assess the effect of poorly constrained OWS efficiency under high-OWS model conditions on the model's results. For this, while sweeping across erosion rates, we varied the parameter "Fraction of non-oxidized Sulfide during Surface erosion" two orders of magnitude above and below its standard model value of 0.01%. In the SI3 plots, we observe that such a four order of magnitude change of one parameter only visibly affects steady-state MS mass, which means that the uncertainty regarding this parameter does not affect other steady-state reservoir masses uncertainties. As for MS, increasing OWS efficiency by two orders of magnitude leads to a decrease of MS $S_{red}$ mass of about one order of magnitude whereas decreasing OWS efficiency by two orders of magnitude leads to an increase in MS $S_{red}$ mass by about two orders of magnitude (see figure SI3g). This would still result in the MS $S_{red}$ mass being two orders of magnitude below present-day values.

As the main transient reservoir that enables fast transfer of sulfur in both redox species between many of the larger reservoirs, oceans and parameters governing oceanic processes play a key role in this model. Hence, testing the model's sensitivity to these parameters enables a better understanding of how oceanic mass fluxes are shaped by several geochemical processes at the same time. We began by testing the effect of baryte precipitation and ocean saturation by varying the baryte sulfate precipitation fraction (the mass fraction of oceanic sulfate removed by baryte precipitation per timestep) for several values of the solubility limit of sulfate in the ocean (see figures SI4). These plots, namely that of oceanic $S_{ox}$, show the interplay between the two regimes of oceanic S cycling in this phase space (see figure SI4j). For the standard ocean $S_{ox}$ saturation limit of 0.03 M, for low baryte precipitation rates (within one order of magnitude of the present-day rate estimate), the oceanic mass of $S_{ox}$ seems to be independent of baryte precipitation rates. This is an indication that the ocean is at its maximum carrying capacity of $S_{ox}$, and that background precipitation (due to excess input of weathered $S_{ox}$) dominates the ocean sulfur cycle. However, beyond a critical baryte removal rate, the baryte precipitation mechanism becomes dominant, and starts removing enough oceanic $S_{ox}$ to stop the ocean from being saturated in $S_{ox}$, decreasing the steady-state ocean $S_{ox}$ mass. This critical point appears to occur at distinct baryte removal rates for distinct maximum oceanic solubilities. The smaller the ocean's carrying capacity of $S_{ox}$ (which implies a more effective normal precipitation mechanism), the more intense the baryte precipitation rates should be in order for the ocean to stop being saturated in $S_{ox}$.

A similar signature to this is found in the majority of the reservoir's SI4 plots, which implies a tight coupling of much larger reservoirs' sulfur budget to the oceanic sulfur cycle. For instance, $S_{ox}$ in MS shows a mirrored trend with respect to the Oce reservoirs, as larger baryte precipitation rates and lower oceanic maximum solubilities lead to steady-state $S_{ox}$ masses in MS up to 30% higher than for the standard model configuration (see figure SI4h). This trend is reflected in mantle and CC reservoirs (see figures SI4a, SI4b and SI4f), albeit with a lower sensitivity (below 1% for mantle reservoirs and below 50% for CC $S_{ox}$). As for the OC $S_{ox}$ reservoir, a tight coupling with the ocean's sulfur cycle is evident, as higher oceanic $S_{ox}$ masses (caused by higher oceanic carrying capacities and lower baryte precipitation rates) allow higher hydrothermal removal of oceanic $S_{ox}$ into the OC, leading to an increase of up to a factor of four in OC $S_{ox}$ mass for one order of magnitude higher oceanic carrying capacities.

We have assessed the effect of this hydrothermal circulation removal of oceanic $S_{ox}$ into OC on the plots of SI5, by considering the same baryte removal rate sweeping for net hydrothermal circulation rates two orders of magnitude above and below the standard model estimate. The HT removal mechanism works by removing a fraction of the ocean Sox every timestep, at a rate that decreases exponentially by a factor of 3 over the 4.5 Gyr of simulation. In the compared cases of SI5, the initial and final removal rates are multiplied by the same factor, keeping the same functional dependence over time. While the same transition between the previously described regimes (dominated by precipitation vs baryte removal) is observed at the same critical baryte removal rate (at a removal rate about two orders of magnitude higher than our best guess for present-day Earth), hydrothermal circulation rates also appear to have an impact on $S_{ox}$ oceanic content (see figure SI5j). Decreasing the hydrothermal circulation rates by two orders of magnitude with respect to our best guess value does not significantly affect the oceanic $S_{ox}$ mass, as for this regime this reservoir is dominated by precipitation due to the ocean's saturated state. However, increasing hydrothermal circulation rates by two orders of magnitude leads to a decrease of oceanic $S_{ox}$ mass by a factor of 20 below the present-day saturated ocean mass, implying that for this set of parameters, the oceanic sulfur cycle becomes dominated by hydrothermal circulation removal. This distinct regime significantly affects the OC $S_{ox}$ reservoir, as it is the sink of the hydrothermally removed ocean sulfate (see Figure SI5d), allowing it to increase its maximum $S_{ox}$ mass by a factor of four with respect to the standard model configuration. Similarly to the oceans, MS becomes depleted in $S_{ox}$ with respect to PEL once HT circulation overrides precipitation as the main sink of oceanic sulfur.

Finally, we also assessed the impact of volcanic outgassing efficiency on the overall abiotic S cycle by sweeping Arc Volcanism Outgas Fraction rates (*i.e.*, the fraction of Arc Volcanism mass flux that is outgassed as $S_{Ox}$ into the atmosphere), from close to 0% to 100% efficiency. This set of parameter sweeping simulations were conducted for both the no-OWS and the high-OWS scenarios, these results are displayed in figure SI6. As expected, increasing the efficiency of arc volcanism outgassing leads to a linear increase of Atmospheric $S_{ox}$ content (Figure SI6k). The tight coupling between atmospheric, oceanic and MS reservoirs (through fast rainout and subsequent precipitation) leads to a similar increase of $S_{ox}$ in MS as a function of Arc Volcanism outgassing (Figure SI6h). This is not directly reflected in the oceanic $S_{ox}$ content (Figure SI6j) as this reservoir is saturated under these conditions, and hence, all input $S_{ox}$ is precipitated to MS. Increased MS $S_{ox}$ content also leads to more rapid S burial through subduction, increasing Mantle reservoir S content (Figures SI6a and SI6b). The OC and CC reservoirs, however, become depleted in S with the increase of $S_{ox}$ volcanic outgas efficiency, as less solid sulfur is produced through volcanism (see Figures SI6c and SI6e).

It is worth mentioning that volcanic outgassing is the second main source of $S_{ox}$ to the MS, after OWS. It is therefore interesting that in shutting down OWS (in the no-OWS scenario) and varying the Volcanic Outgas efficiency from 100% to 0% (see plot in Figure SI6h) there is an 80% decrease in the MS $S_{ox}$ mass. Hence, even with its two main sources of $S_{ox}$ (OWS and Volcanic Outgas) shut down in this model, the MS $S_{ox}$ mass never falls below four times the current PEL of MS $S_{ox}$ even in these extreme scenarios, while the MS $S_{red}$ mass is still orders of magnitude below PEL. A possible explanation for this is the lack of biological activity in our models. In particular, that of microbial sulfate reduction (MSR) in sedimentary settings (Habicht & Canfield, 1997; Wasmund *et al.* 2017; Jorgenssen *et al.,* 2019). This is in agreement with the idea that microbiota (especially in MS environments) play a key role in Earth's S-Cycle, by reducing large fractions of MS $S_{ox}$. Current MS MSR is estimated at $1.1 \times 10^{13}$ moles S per year, or roughly $3.6 \times 10^{11}$ kg S (Bowles *et al.*, 2014), which would be sufficient to reduce the modern oceanic $S_{ox}$ reservoir in ~$3.5 \times 10^6$ years, or the current MS $S_{ox}$ reservoir in a roughly geologically comparable ~$6 \times 10^5$ years.

## 4 – Conclusions

Modeling long-term sulfur cycling on terrestrial planets with dynamic hydrological cycles is complex but may allow for easy identification of where the emergence of biology has significantly intervened in distributing remotely detectable sulfur species among planetary reservoirs, and in determining the redox state of sulfur in those reservoirs. That such crude models can also effectively match modern terrestrial sulfur dynamics suggests general principles of sulfur cycling can also be applied to exoplanets. This model also suggests that abiological sulfur cycling on rocky planets over geological timescales requires careful consideration to disentangle contributions from atmospheric and hydrologic processes, both of which depend on initial sulfur inventories, insolation, and weathering processes, which have been significantly steered by biological processes on Earth since the origin of life.

The movement of redox-active elements such as sulfur (and carbon and nitrogen) also depends on the rates of transport and transformation of the redox states of other especially non-volatile, elements such as transition metals (especially iron, due to its cosmic abundance). In this respect, oxygen and sulfur can be seen as fast as slow mediators of electron transport, respectively. The saturation of various reservoirs and transport processes for sulfur are also important for the global chemodynamics of sulfur. For example, the aqueous behavior of reduced sulfur species in the presence of transition metals or of oxidized sulfur in the presence of alkaline earth metals such as calcium, often slow their hydrologic transport, such that the amount of water relative to a planet's mass may significantly impact planetary surface sulfur dynamics.

What biology seems to be especially adept at accomplishing with sulfur is the rapid and novel — with respect to abiological processes — mobilization of sulfur species between geochemical reservoirs in two fundamental ways: first by geologically rapid disproportionation of sulfur in sediments, and second by the acceleration of sorting processes, *e.g.* increasing surface sulfur weathering rates via the production of $O_2$. One could expect that exoplanets with remotely measurable $SO_2$ or $H_2S$ abundances do not have surface oceans, since it is likely these species would be sequestered into oceans, due to the Henry's Law-governed partitioning of these species (Kaltenegger and Sasselov, 2009; Loftus *et al*, 2019). Even early endowment of a planet with more sulfur than Earth would end up with more of it sequestered in non-atmospheric sulfur-accepting reservoirs. Along the same lines, it has recently been suggested that $CO_2$ may be an anti-biosignature (Triaud *et al.*, 2023), as liquid surface water, which may be required for the origins of life, should generally make $CO_2$ a minor component of the atmospheres of planets harboring liquid surface oceans.

While some lower oxidation state sulfur-containing gasses have been suggested to be potential exoplanetary biosignatures (Domagal-Goldman *et al.*, 2011; Schwieterman *et al.*, 2018; Madhusudhan *et al.*, 2023, 2025), as with other proposed atmospheric biosignatures, interpretation requires considering abiotic production pathways and potential false positives in the context of the planet's environment (Harman *et al.*, 2018). Moreover, lower oxidation state sulfur species are generally not photochemically stable, especially in the presence of water vapor, being prone to disproportionation and oxidation in the presence of OH radicals (Barnes *et al.,* 1986; Tyndall & Ravishankara 1991; Domagal-Goldman *et al.,*

2011). Thus, high atmospheric abundances of low-redox state sulfur may be difficult to maintain on planets harboring liquid water oceans, while high atmospheric abundances of high-redox state sulfur may also be difficult to maintain on ocean worlds due to the photochemical conversion of $SO_2$ to highly soluble sulfate. Thus, high S-species atmospheres may instead be hallmarks of planets lacking geodynamic processes, and especially oceans, which remove sulfur from the atmosphere, *e.g.* Venus-like planets (Kaltenegger & Sasselov, 2009).

## 5 - Acknowledgements

The authors would like to thank the BMSIS YSP program for programmatic support. This work was supported by Fundação para a Ciência e Tecnologia (FCT) of reference PTDC/FIS-AST/29942/2017, through national funds and by FEDER through COMPETE 2020 of reference POCI-01-0145-FEDER-007672, and through the research grants UIDB/04434/2020, UIDP/04434/2020 and UID/04434/2025. RRS acknowledges funding through the FCT fellowship grant 2024.02527.BD.

## 6 - Author Contributions

Conceptualization: RRS, JAM, MJM, HJC; Formal analysis; RRS, JAM, HJC; Investigation; RRS, JAM, HJC; Methodology; RRS, JAM, HJC; Project administration; HJC; Software; RRS, JAM, MJM, HJ, HJC; Supervision; HJC; Validation; RRS, JAM, HJC; Visualization; RRS, JAM; Roles/Writing - original draft; RRS, JAM, HJC; Writing - review & editing: RRS, JAM, HJC, MAP;

**7 - Code availability**: The code for this model can be found on GitHub at
*https://github.com/Javi786/Abiotic-S-cycling-model*

## 8 - References

**Tables:**

**Table 1.** Estimated sulfur reservoirs on Venus, Earth and Mars. 1. https://nssdc.gsfc.nasa.gov/planetary/factsheet/  2. Estimated using Wedepohl (1984) meteoritic feedstock estimate (~2 % of planetary mass, see below) 3. Estimated using Dreibus & Palme 1996 with a maximum bulk Earth S-content of 0.56% 4. see Franz *et al.,* (2019) for a good discussion. NA = not applicable.

| Planet | Planet Mass (kg)[1] | Estimated Planetary S mass (kg)[2] | Surface Atmospheric Pressure (bar) | Atmosphere Mass (kg) | Atmosphere S Mass (kg) | Oceanic S Mass (kg) |
|---|---|---|---|---|---|---|
| Venus | $4.87 \times 10^{24}$ | $1.00 \times 10^{23}$ | 92 | $4.8 \times 10^{20}$ | $7.2 \times 10^{16}$ | NA |
| Earth | $5.97 \times 10^{24}$ | $3.34 \times 10^{22}$ [3] | 1.014 | $5.15 \times 10^{18}$ | $1 - 4.8 \times 10^{9}$ | $1.4 \times 10^{18}$ |
| Mars[4] | $6.42 \times 10^{23}$ | $1.31 \times 10^{22}$ | $4 - 8.7 \times 10^{-3}$ | $2.5 \times 10^{16}$ | $5.7 - 7.8 \times 10^{8}$ | NA |

**Table 2.** Estimated modern-day S-content in terrestrial reservoirs. Some reservoirs contain both $S_{ox}$ and $S_{red}$, whereas some reservoirs only contain one of the two. Values are rounded to two significant figures. a. Dreibus & Palme (1996), b. Wedepohl (1984) estimated as 500 ppm, c. Wedepohl (1984), d. Holser (1988), e. Wedepohl (1984) (sediments S content ~4250 ppm), Goldhaber (2003) (suggested >80% of S in marine sediments is pyrite), f. Estimated using the values of 0.1 to 1.1 nM of Cutter and Krahforst (1988), g. (Martin 2016), h. Papuc & Davies (2008), i. Lodders (1998), j. Nolet (2011), k. Jusino-Maldonado *et al.* (2022), l. Peterson & Depaolo (2007), m. Stern & Scholl 2010, n. Peterson & Depaolo (2007), o. Wedepohl (1995), p. Hay *et al.*, (1988), q. Henderson & Henderson (2009).

| Reservoir | Estimated modern reservoir $S_{ox}$ or $S_{red}$ content (kg S) | | Modern reservoir mass (kg) |
|---|---|---|---|
| | $S_{red}$ | $S_{ox}$ | |
| **Core** | 3.4 x $10^{22}$ [a] | — | 2.0 x $10^{24}$ [h] |
| **Lower Mantle** | 1.5 x $10^{21}$ [b] | — | 3.0 x $10^{24}$ [i] |
| **Upper Mantle** | 5.5 x $10^{20}$ [c] | — | 1.1 x $10^{24}$ [j] |
| **Oceanic crust** | 3.4 x $10^{18}$ [d] | 4.8 x $10^{17}$ [d] | 9.0 x $10^{21}$ [k,l] |
| **Continental crust** | 1.8 x $10^{19}$ [d] | 6.2 x $10^{18}$ [d] | 1.9 x $10^{22}$ [m] - 2.2 x $10^{22}$ [n]<br><br>1.6 x $10^{22}$ [o] |
| **Marine sediments** | 8.8 x $10^{17}$ [e] | 2.2 x $10^{17}$ [e] | 2.6 x $10^{20}$ [p] |
| **Oceans** | 4.4 x $10^{9}$ - 4.9 x $10^{10}$ [f] | 1.3 x $10^{18}$ [g] | 1.4 x $10^{21}$ [q] |
| **Atmosphere** | — | 4.8 x $10^{9}$ [g] | 5.1 x $10^{18}$ [q] |
| **Total S Excluding Core** | 2.1 x $10^{21}$ | 8.2 x $10^{18}$ | - |
| **Total S Including Core** | 3.61 x $10^{22}$ | | - |

**Table 3.** Estimated values for the main parameters used in this model. See methods section for detailed discussion of parameter selection. Parameters varied for parameter sensitivity analysis are marked with *.

| Model Parameter | Parameter Value | Parameter Units |
|---|---|---|

| Surface erosion Rate *, α | 2.2 x 10⁻⁴ for high-OWS conditions | 2.2 x 10⁻⁸ for no-OWS conditions | m/year |
|---|---|---|---|
| Fraction of non-oxidized Sulfide during Surface erosion * | 0.01% for high-OWS conditions | 100% for no-OWS conditions | - |
| Atmospheric Half-Life of $SO_2$ ($\tau[SO_2]$) | 0.25 | | year |
| Solubility Limit of Sulfate in Ocean * | $3.0 \times 10^{-2}$ | | mol of $S_{ox}$/L |
| Solubility Limit of Sulfide in Ocean | $1.0 \times 10^{-9}$ | | Mol of $S_{red}$/L |
| Ocean Volume | $1.35 \times 10^{21}$ | | L |
| Baryte Sulfate Precipitation Fraction * | $10^{-6}$ | | Fraction/time step |
| Subduction Timescale *, D | 100 | | Ma |
| Accretion Efficiency, ε | 0.05 | | - |
| Arc Volcanism (associated to subduction): Fraction of OC/MS S mass entering the subduction process that is excreted by Arc Volcanism: φ | 0.2 | | - |

| | | |
|---|---|---|
| Fraction of the S mass being excreted by Arc Volcanism that is emitted as outgassing (SO$_2$): | 0.5 | - |
| MORB Volcanic Output Rate: | 20 | km³ / yr |
| Hotspot Volcanic Output Rate | 2.5 | km³ / yr |
| Fraction of MORB and Hotspot Volcanism S Mass emitted as outgassing SO$_2$: | 0.2 | - |
| Initial volcanic multiplying factor, **f** | 3 | - |
| Volcanic Exponential decay timescale | 150 | Ma |
| **Volcanism S enrichment factors** | | |
| MOR volcanism | 2.0 | - |
| Arc volcanism | 1.7 | - |
| Hotspot volcanism | 2.0 | - |
| **Mantle mixing rates** | | |
| Min value **F$_0$** | 1.0 x 10$^{-8}$ | Yr$^{-1}$ |
| Max value **F$_1$** | 3.0 x 10$^{-8}$ | Yr$^{-1}$ |
| Convection Exponential decay timescale: | 150 | Ma |
| **Hydrothermal Circulation** * | | |
| Min value $F_0$ | $1.0 \times 10^{-8}$ | $Yr^{-1}$ |
| Max value $F_1$ | $3.0 \times 10^{-8}$ | $Yr^{-1}$ |
| HT Exponential decay timescale: | 150 | Ma |
| **ET Input** | | |
| Min value $F_0$ | $2.0 \times 10^{5}$ | kg / yr |
| Max value $F_1$ | $2.0 \times 10^{8}$ | kg / yr |
| ET Input Exponential decay timescale: | 150 | Ma |

**Figures and Figure Legends :**

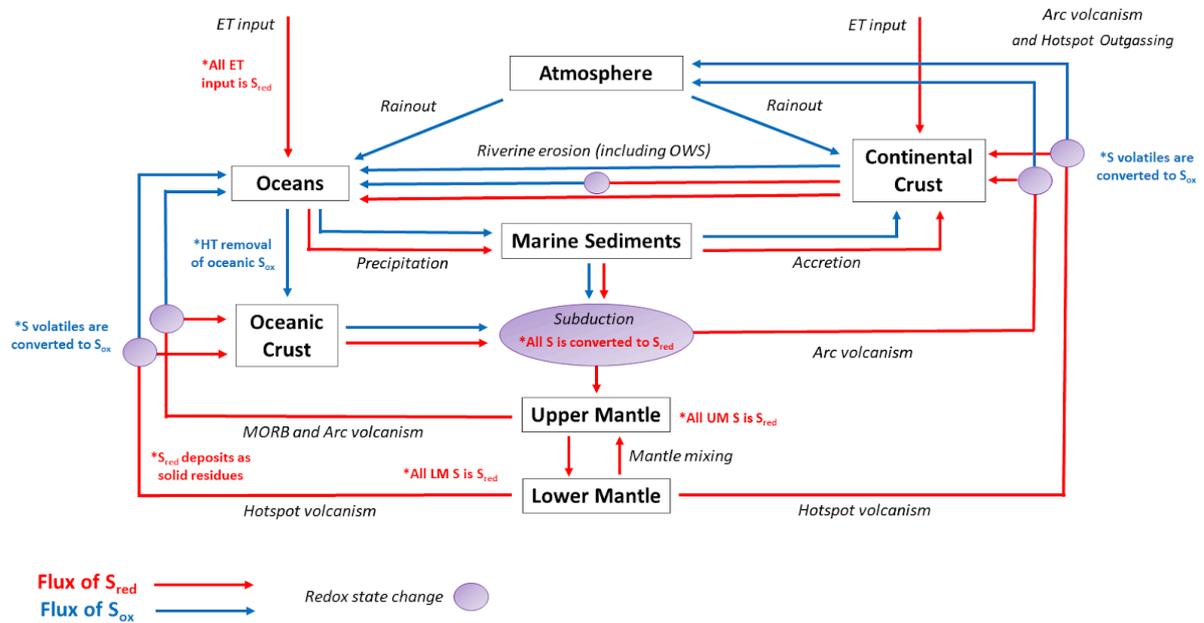

**Figure 1**. Flowchart of the planetary geochemical cycling model of sulfur modeled in this work. In this "box and arrow" model, the boxes correspond to the seven geochemical reservoirs (Lower and Upper Mantle, Oceanic and Continental Crust, Marine Sediments, Oceans and Atmosphere), while arrows represent the distinct sulfur fluxes arising from the geochemical processes included in this model. Processes which change sulfur redox state significantly during processing are marked in purple.

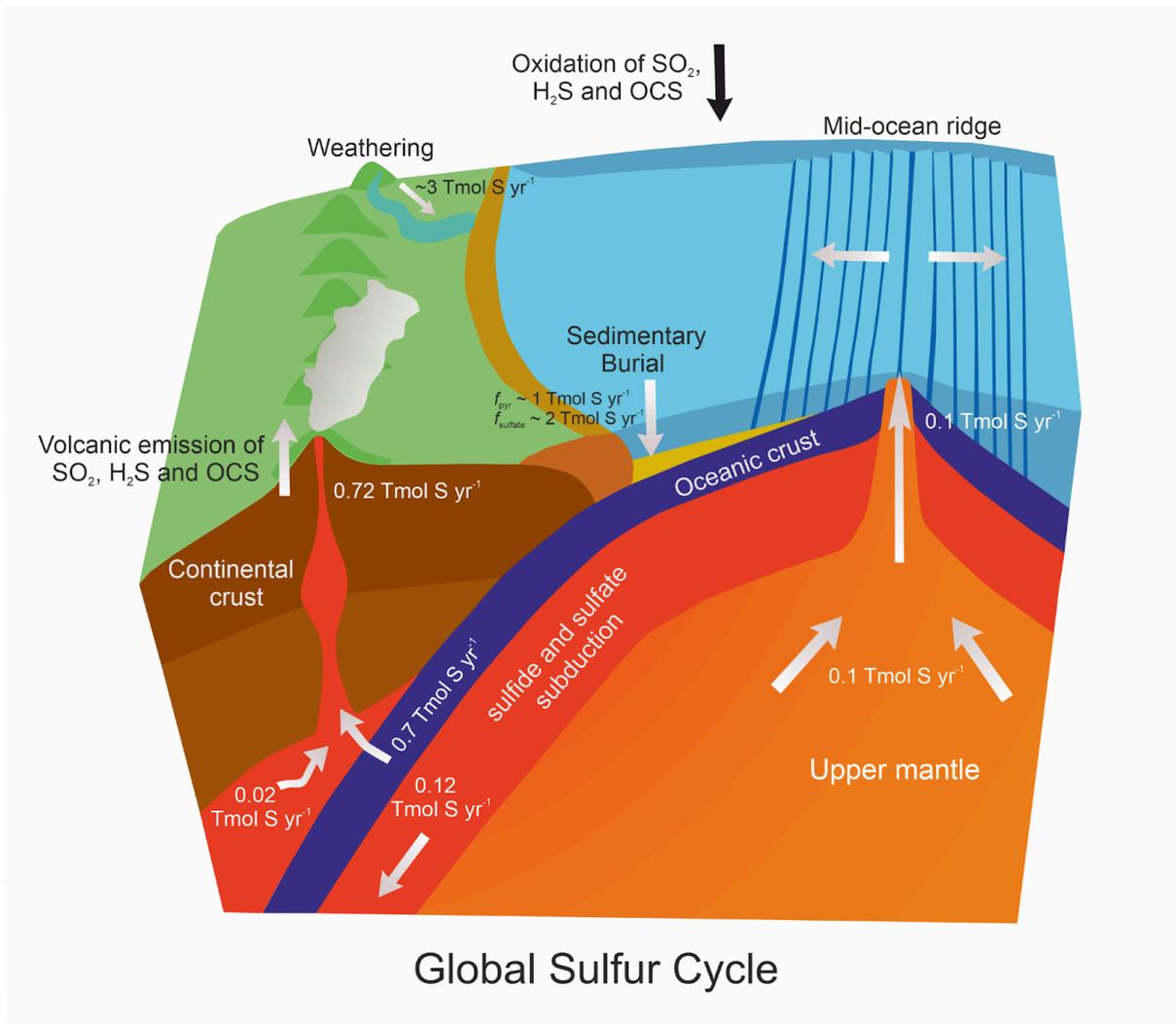

**Figure 2:** Schemetic diagram of global sulfur cycle with major fluxes of sulfur between the deep Earth and surface Earth (in 10$^{12}$ moles of sulfur per year, denoted as Tmol S yr$^{–1}$; based on Kagoshima *et al.,* 2015; Fakhraee *et al.,* 2025).

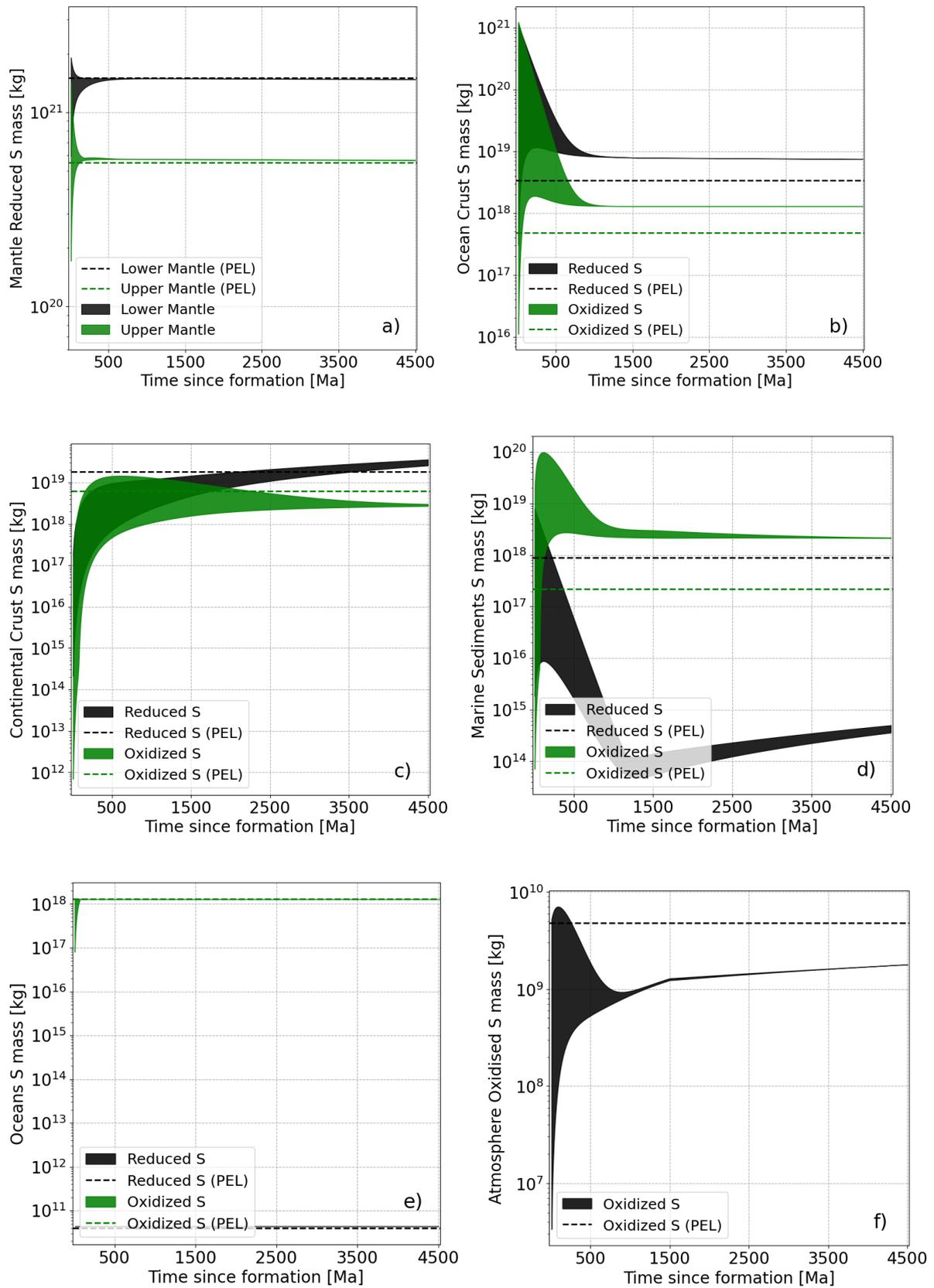

**Figure 3.** Temporal evolution of $S_{ox}$ and $S_{red}$ masses across the model's reservoirs, for the best-guess geochemical model parameters for a no-OWS model configuration, mimicking a pre-GOE Earth, compared to Present Earth Levels (PEL), the best estimations for the S

budgets of these reservoirs in present Earth. These plots results are discussed in detail in section 3.1, addressing possible explanations for the differences observed between steady state reservoir mass values and PEL. Subplots: (a) - Lower Mantle - LM - and Upper Mantle - UM; (b) - Oceanic Crust - OC; (c) - Continental Crust - CC; (d) -Marine Sediments - MS; (e) - Oceans - Oce; and (f) - Atmosphere – Atm. Comparison with Present Earth Level (PEL) of each reservoir's Reduced and Oxidized Sulfur content.

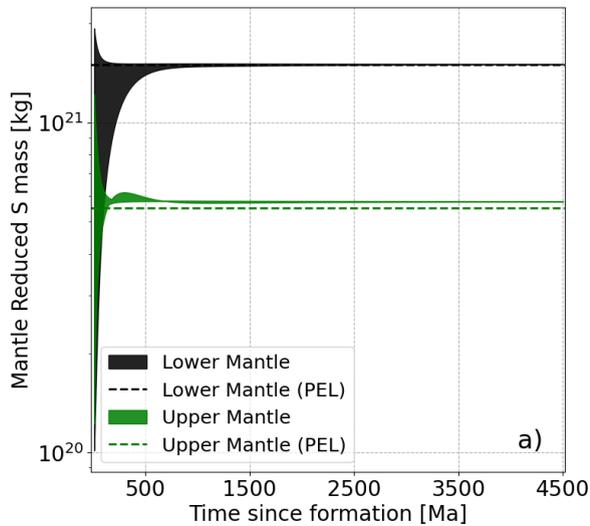
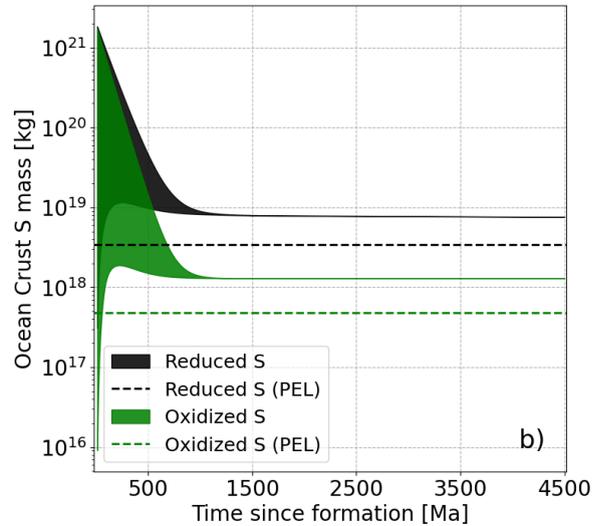
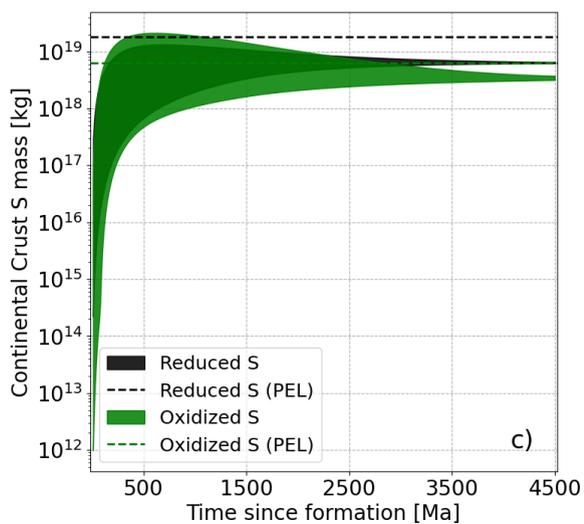
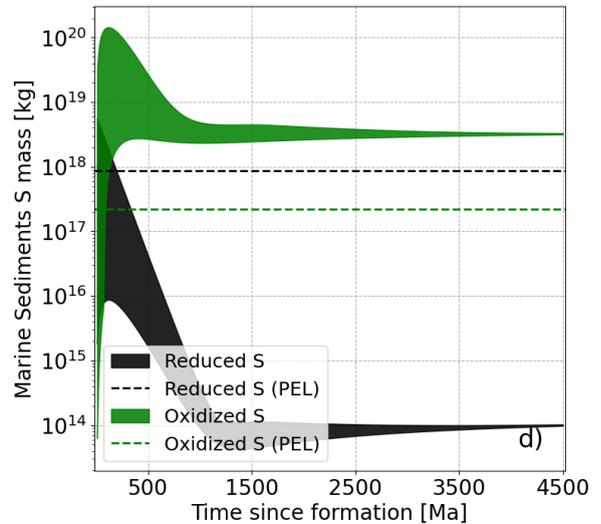

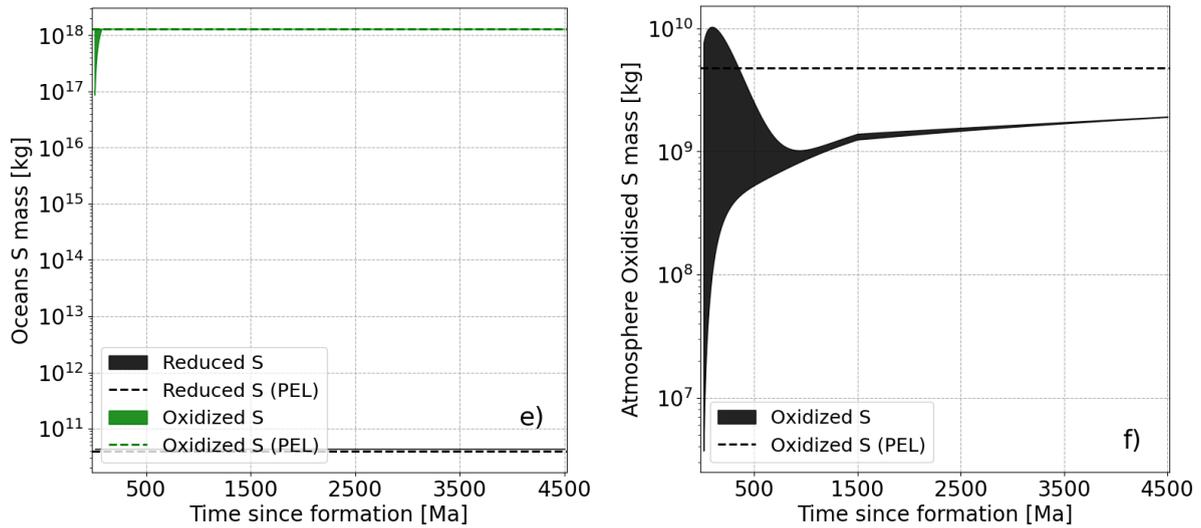

**Figure 4.** Temporal evolution of $S_{ox}$ and $S_{red}$ masses across the model's reservoirs, for the best-guess initial sulfur inventory model parameters for the high-OWS model configuration, mimicking a post-GOE Earth, compared to Present Earth Levels (PEL). These plots results are discussed in detail in section 3.1, addressing possible explanations for the differences observed between steady state reservoir mass values and PEL. Subplots: (a) - Lower Mantle - LM - and Upper Mantle - UM; (b) - Oceanic Crust - OC; (c) - Continental Crust - CC; (d) -Marine Sediments - MS; (e) - Oceans - Oce; and (f) - Atmosphere - Atm. Comparison with Present Earth Level (PEL) of each reservoir's Reduced and Oxidized Sulfur content.

# Supplementary Information for "Global Abiotic Sulfur Cycling on Earth-like Terrestrial Planets"

Rafael Riança-Silva[1, 2, 3, 5] *, Javed Akhter Mondal[4, 5] *, Matthew A. Pasek[6], Henry Jurney[5], Marcos Jusino-Maldonado[5, 7], Henderson James Cleaves II[5, 8, 9, 10]

**Affiliations**

1. Instituto de Astrofísica e Ciências do Espaço, Observatório Astronómico de Lisboa, Ed. Leste, Tapada da Ajuda, 1349-018 Lisbon, Portugal
2. Departamento de Física, Faculdade de Ciências, Universidade de Lisboa, 1749-016 Lisboa, Portugal
3. Department of Physics and Astronomy, University College London, Gower Street, WC1E 6BT London, United Kingdom
4. Deep Space Exploration Laboratory/CAS Key Laboratory of Crust-Mantle Materials and Environments, University of Science and Technology of China, Hefei 230026, China
5. Blue Marble Space Institute of Science, Seattle, USA
6. Earth and Environmental Science, Rensselaer Polytechnic Institute, Troy NY, USA
7. Planetary Habitability Laboratory, University of Puerto Rico at Arecibo, Puerto Rico
8. Earth-Life Science Institute, Tokyo Institute of Technology, Tokyo, Japan
9. Earth and Planets Laboratory, Carnegie Institution of Washington, Washington DC, USA
10. Department of Chemistry, Howard University, Washington DC, USA

\* Authors contributed equally for this work

Corresponding author email: rdsilva@fc.ul.pt

The code for this model can be found on GitHub at *https://github.com/Javi786/Abiotic-S-cycling-model*

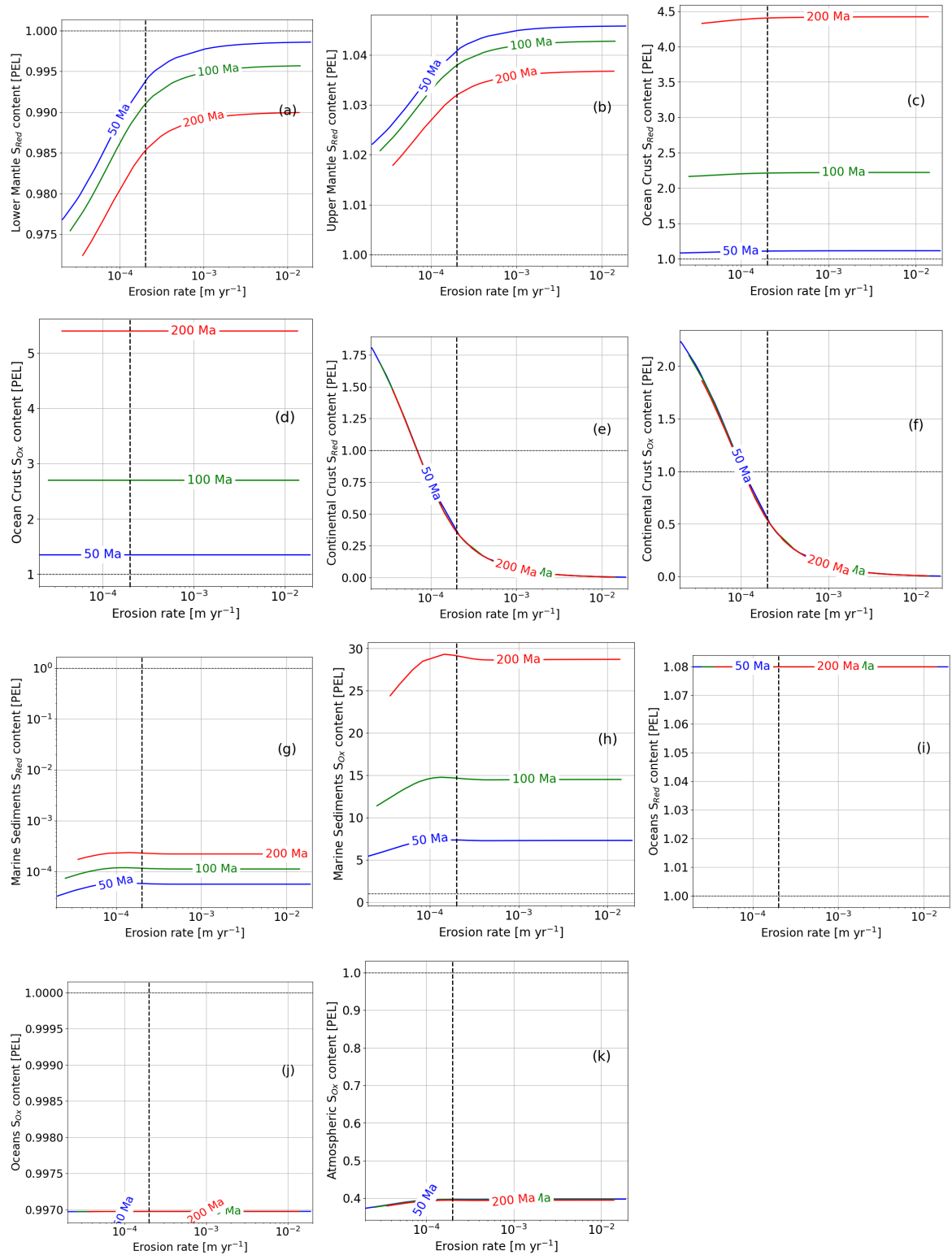

**Figure SI1.** Varying erosion rates for several OC recycling timescales (50 Ma, 100 Ma and 200 Ma).

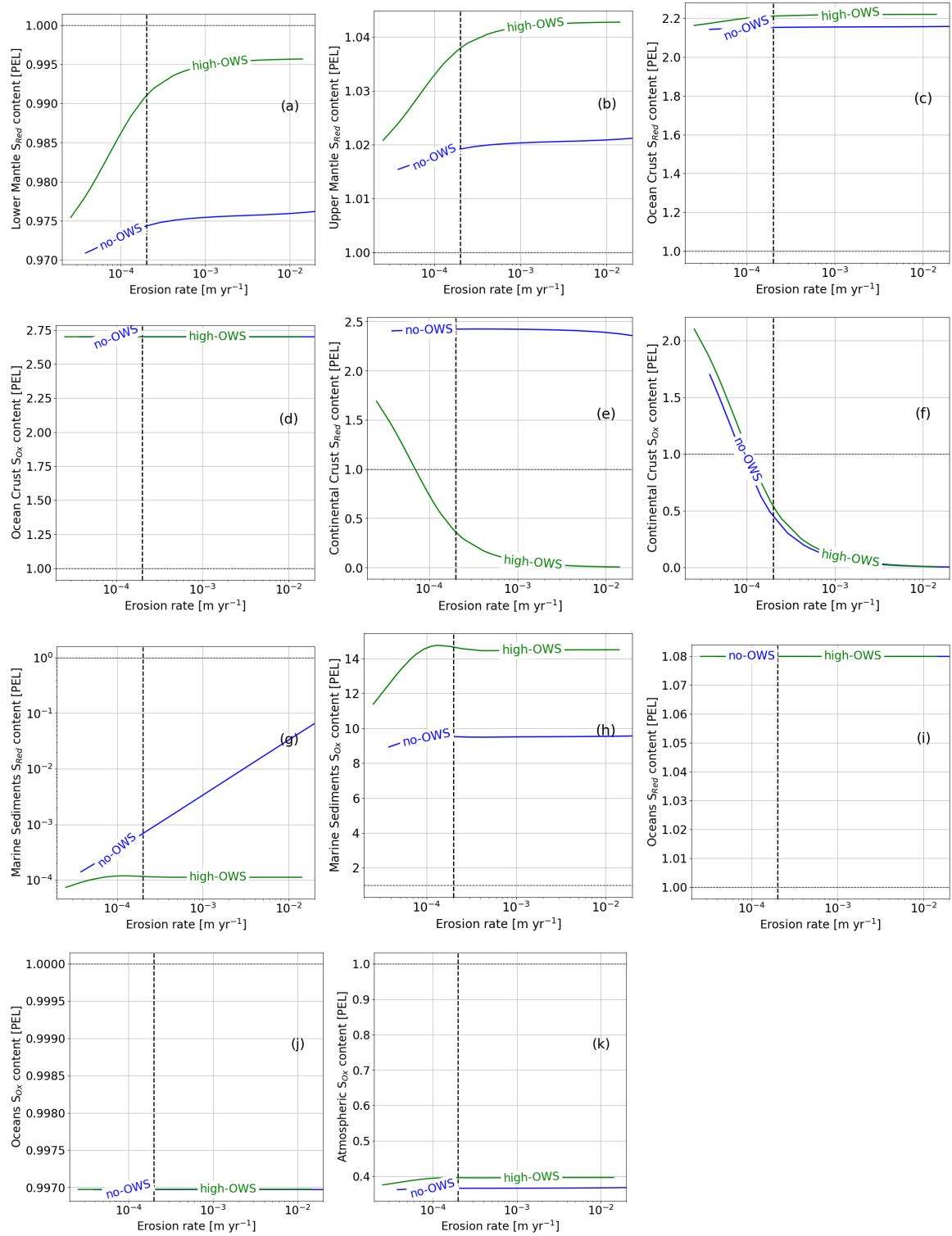

**Figure SI2.** Varying erosion rates for no-OWS and high-OWS model configurations (see GOE section of the main text).

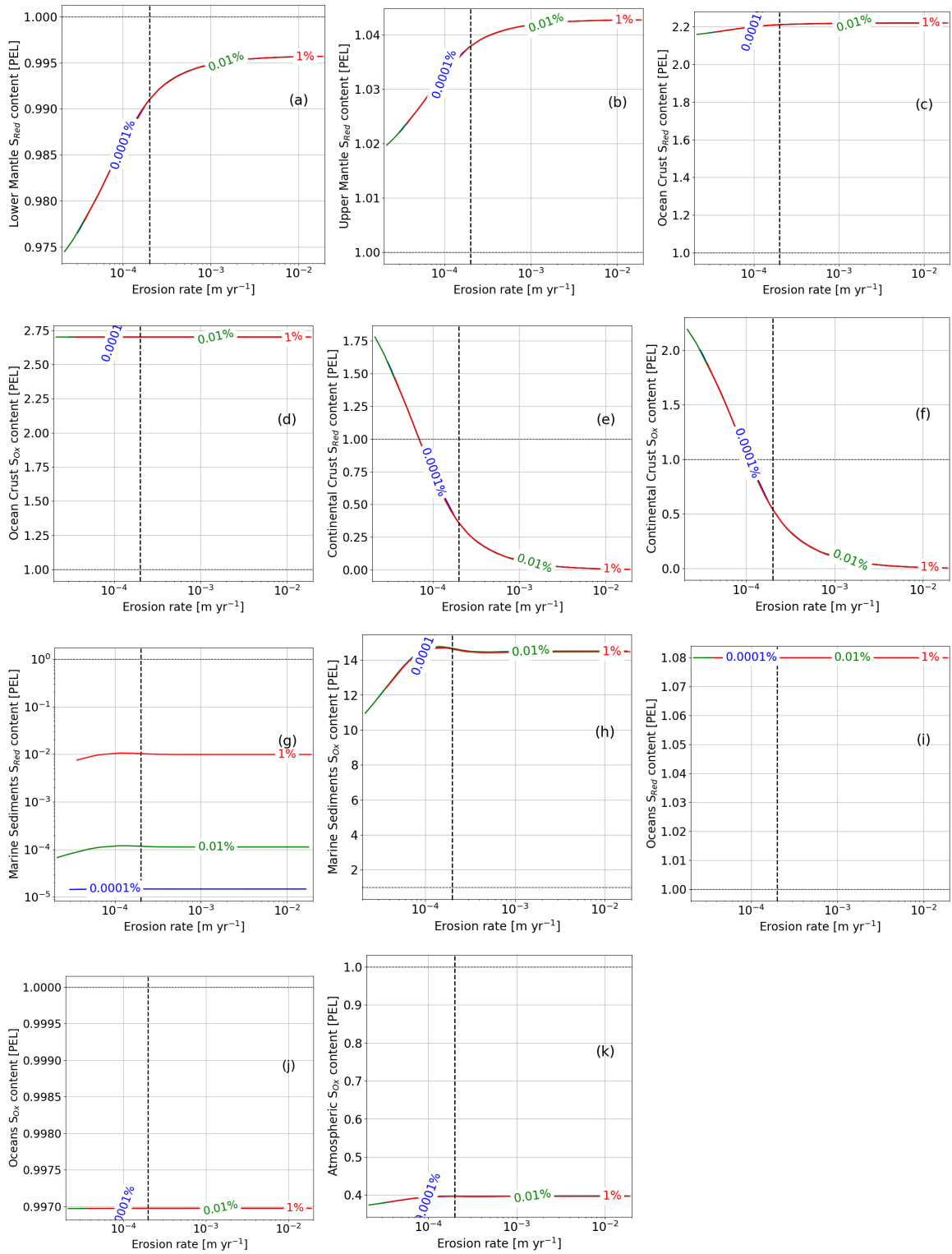

**Figure SI3.** Varying erosion rates for several fractions of non-oxidized Sulfide during OWS (the fraction of S$_{red}$ reaching the ocean during the process of OWS, where the vast majority of S$_{red}$ is oxidized).

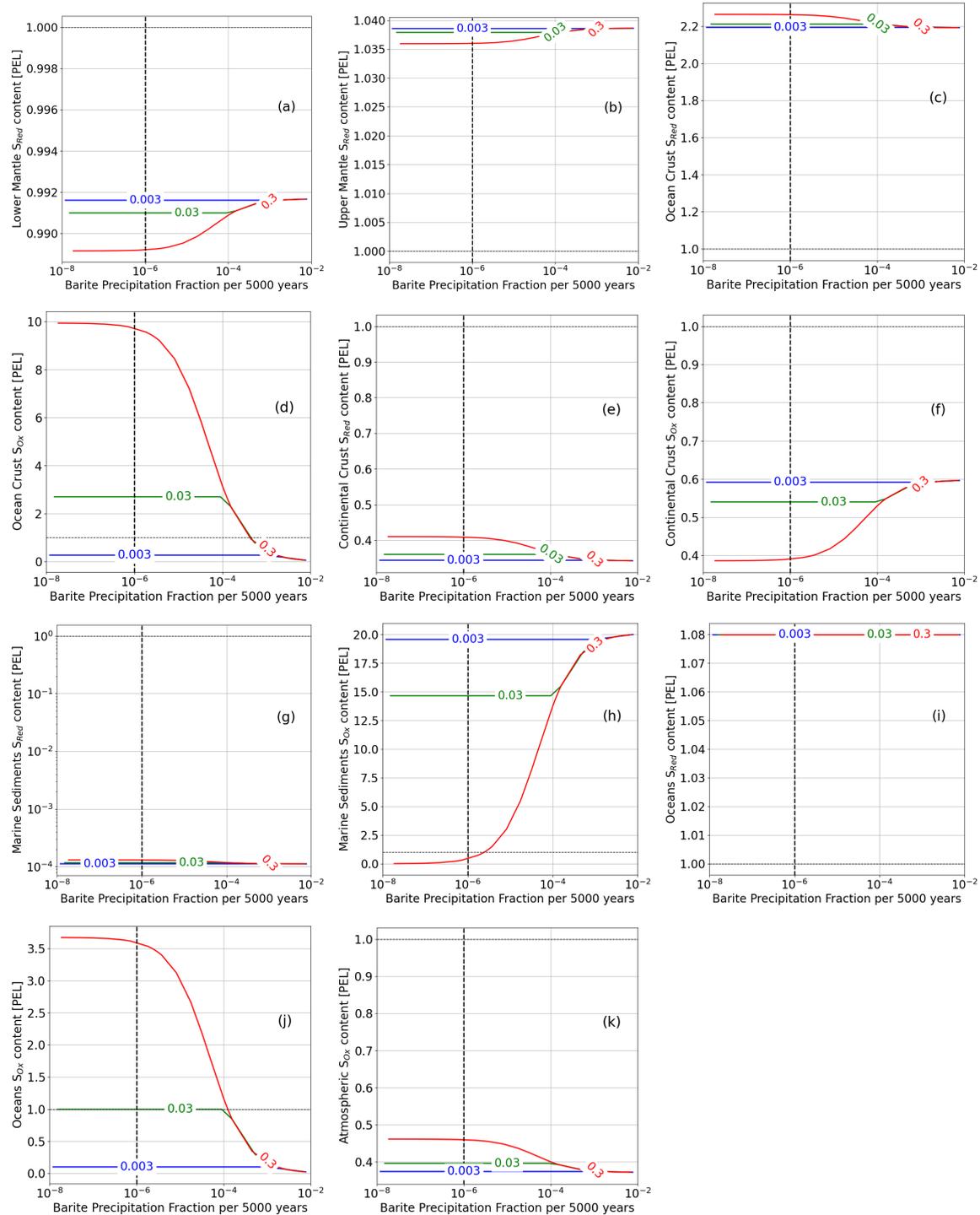

**Figure SI4.** Varying baryte precipitation rates (fraction of oceanic $S_{ox}$ mass per timescale) for several oceanic maximum $S_{ox}$ solubilities (in M).

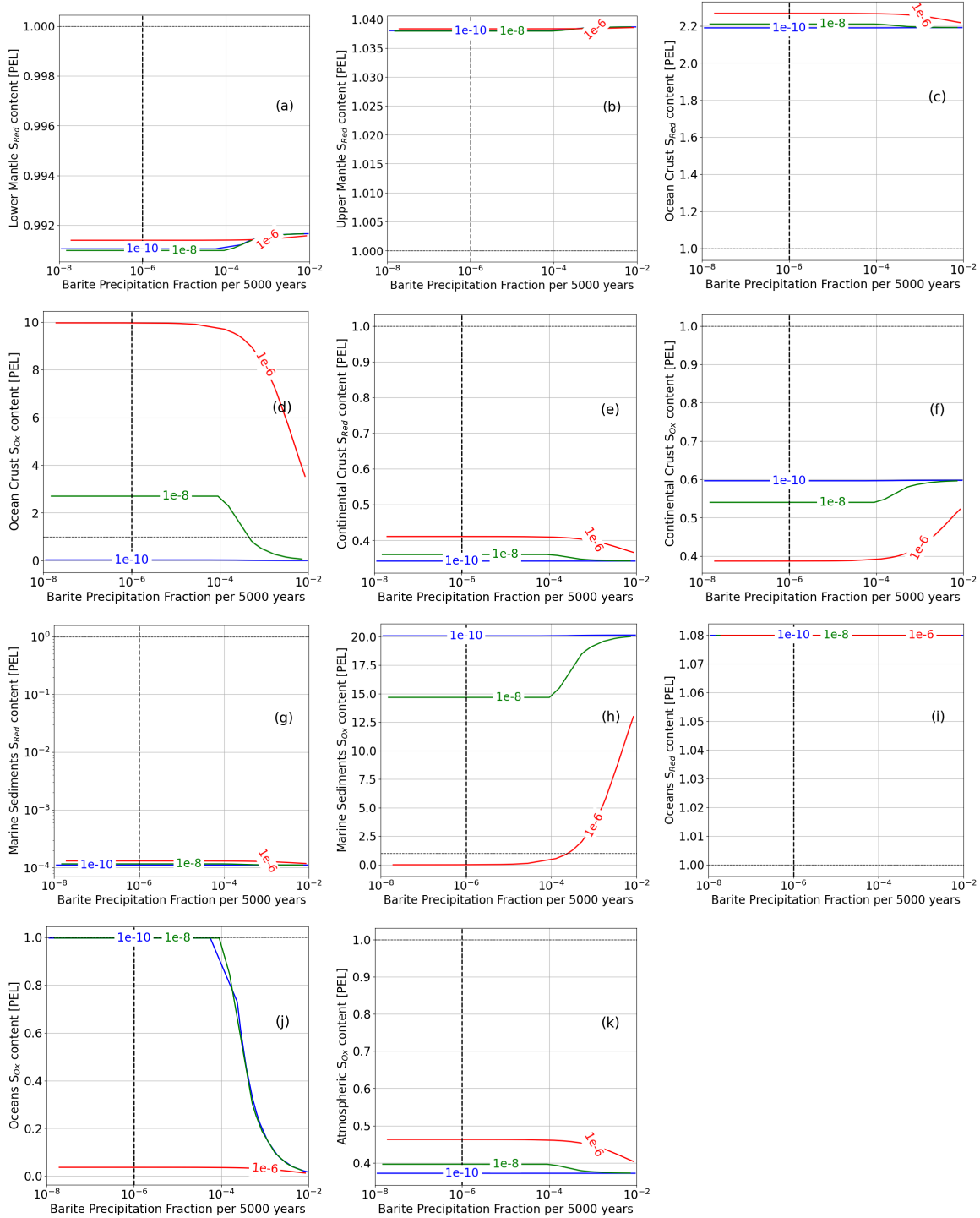

**Figure SI5.** Varying baryte precipitation rates (fraction of oceanic $S_{ox}$ mass per timescale) for several net hydrothermal circulation rates (in ocean volumes per year).

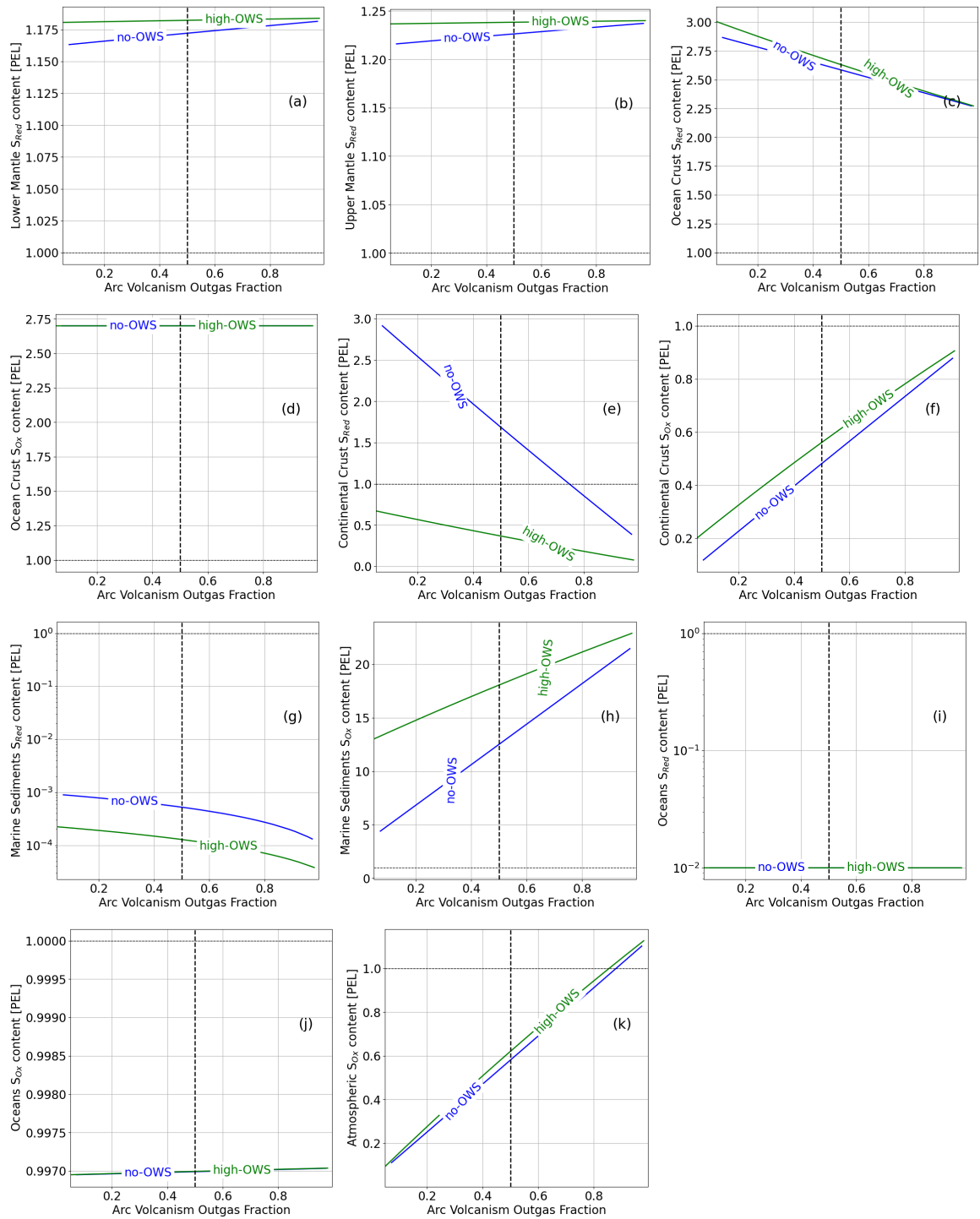

**Figure SI6.** Varying arc volcanism outgas fraction rates (fraction of arc volcanism mass flux that is outgassed as $S_{Ox}$ into the atmosphere) for no-OWS and high-OWS model configurations (see GOE section of the main text).